%% file: paper-marketing.tex
\documentclass[11pt]{article}

\usepackage[utf8]{inputenc}
\usepackage{amsmath, amssymb}
\usepackage{graphicx}
\usepackage{float}
\usepackage{booktabs}
\usepackage{natbib}
\usepackage{changes}
\usepackage{geometry}
\usepackage{setspace}
\usepackage{subcaption}
\usepackage[hyphens]{url}   
\usepackage{hyperref}

\title{Deceptively Framed Lotteries in Consumer Markets\thanks{We are grateful to Simon Cordes, Vincent Eulenberg, Alexander Koch, Hannah Schildberg-Hörisch, Leon Landes, Anna Ressi, and Florian Zimmermann for helpful comments. We also wish to thank conference and seminar audiences at the MPI for Research on Collective Goods, the DICE Winter School, and the Workshop of the Research Group on Consumer Preferences, Consumer Mistakes, and Firms’ Response. We gratefully acknowledge financial support by the German Research Foundation (DFG project No. 462020252).}
}
\author{
    Markus Dertwinkel-Kalt\thanks{University of Münster and Max Planck Institute for Research on Collective Goods (Bonn). Email: markus.dertwinkel-kalt@wiwi.uni-muenster.de.} \and
    Hans-Theo Normann\thanks{Düsseldorf Institute for Competition Economics (DICE) and Max Planck Institute for Research on Collective Goods (Bonn). Email: normann@hhu.de.} \and
    Jan-Niklas Tiede\thanks{Düsseldorf Institute for Competition Economics (DICE). Email: tiede@dice.hhu.de.} \and
    Tobias Werner\thanks{University of Southampton and Center for Humans and Machines at the Max Planck Institute for Human Development. Email: tobias.felix.werner@soton.ac.uk.}
}
\date{\today}

\begin{document}
\onehalfspacing

\maketitle

\thispagestyle{empty}

\pagenumbering{roman}
\setcounter{page}{1}

\begin{abstract} \noindent 
Consumers often face products sold as lotteries rather than fixed outcomes. A prominent case is the loot box in video games, where players pay for randomized rewards. We investigate how presentation formats shape consumer beliefs and willingness to pay. In an online experiment with 802 participants, sellers could frame lotteries using two common manipulations: censoring outcome probabilities and selectively highlighting rare successes. More than 80\% of sellers adopted such deceptive frames, particularly when both manipulations were available. These choices substantially inflated buyer beliefs and increased willingness to pay of up to six times the expected value. Sellers anticipated this effect and raised prices accordingly. Our results show how deceptive framing systematically shifts consumer beliefs and enables firms to extract additional surplus. For marketing practice, this highlights the strategic value of framing tools in probabilistic selling models; for policy, it underscores the importance of transparency requirements in protecting consumers.
\end{abstract}



\newpage
\pagenumbering{arabic}
\setcounter{page}{1}

\input{Introduction-Marketing}
\input{Related_Literature}
\input{Experimental_Design}
\input{Hypotheses}

\input{Results}
\input{Conclusion-Marketing}

\bibliographystyle{apalike}
\bibliography{references}

\newpage
\appendix
\section*{Appendix}
\renewcommand{\thefigure}{A\arabic{figure}}
\renewcommand{\thetable}{A\arabic{table}}
\setcounter{figure}{0}
\setcounter{table}{0}
\input{Appendix_A}

\clearpage
\renewcommand{\thefigure}{B\arabic{figure}}
\renewcommand{\thetable}{B\arabic{table}}
\setcounter{figure}{0}
\setcounter{table}{0}
\input{Appendix_B}
\clearpage
\renewcommand{\thefigure}{C\arabic{figure}}
\renewcommand{\thetable}{C\arabic{table}}
\setcounter{figure}{0}
\setcounter{table}{0}
\input{Appendix_C}

\end{document}

%% file: Introduction-Marketing.tex
\section{Introduction}
\label{sec:Introduction}

In many consumer markets, sellers increasingly design their products not as fixed goods but as lotteries. Instead of purchasing a specific item, consumers pay for a chance to receive one of several possible outcomes, some common and low in value, others rare and highly desirable.  This sales model has proliferated, showing up in a wide range of settings, including trading cards, arcade machines, tickets for soccer matches, online mystery boxes, and collectible toys. What unites these settings is not just the randomized nature of the transaction, but the strategic use of presentation formats that shape consumer beliefs and behavior.

A particularly salient example is loot boxes. With this form of in-game microtransactions, players pay to receive randomized virtual rewards. They are deeply embedded in the monetization strategies of major video games and generate billions in annual revenue.\footnote{The market for online microtransactions has expanded significantly in recent years and is projected to reach \$121.2 billion by 2028 \citep{BRC2024}. By 2020 already, about 60\% of the top-grossing games in both the Google Play Store and Apple App Store incorporated loot boxes \citep{zendle2020prevalence}.} Loot boxes are a lottery-like way to obtain virtual items, such as cosmetic upgrades or items that affect gameplay. They are usually delivered through randomized formats like card packs, prize wheels, or digital crates. Players can purchase them directly or earn them through in-game achievements, but the contents are typically unknown until after the box is opened.

The strategic appeal of such formats lies in their ability to increase the willingness to pay without altering the underlying product. By shaping expectations and highlighting certain outcomes, firms can create perceived value beyond the objective odds. This, however, has sparked controversy. In the case of loot boxes, concerns have been raised that they mimic gambling mechanics, encourage overspending, and disproportionately affect younger or more vulnerable consumers \citep{drummond2018video, king2018predatory}. These concerns stem from design features that systematically distort beliefs: often not through the lottery itself, but through \emph{how the lottery is presented}.

\begin{figure}[!ht]
    \centering
    \caption{Communication of loot box odds in \textit{Apex Legends}} 
    \includegraphics[width=0.8\textwidth]{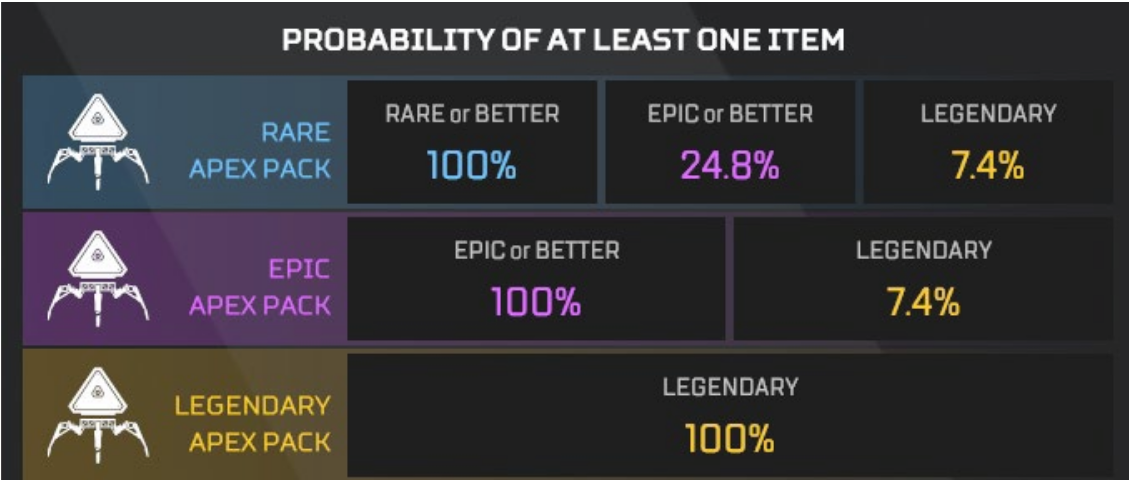}
    \label{fig:Apex Odds}
    \begin{flushleft} 
        \footnotesize
        \setlength{\parskip}{0em} 
        Note: In \textit{Apex Legends} odds are only reported at the tier level.
    \end{flushleft}
\end{figure}

Two mechanisms are especially relevant for marketing practice. The first is censored odds: the true probabilities of receiving different outcomes are often hidden, bundled into vague categories, or disclosed only for low-value items. One video game that utilizes censored odds is \textit{Apex Legends}. Figure~\ref{fig:Apex Odds} displays how probabilities in \textit{Apex Legends} are reported only at the tier level. For lower rarity tiers, the probability is also combined with the probabilities for higher tier items. Other examples for the use of deceptive features in video games can be found in Appendix C.
For instance, in \emph{EA Sports FC}\footnote{\emph{EA Sports FC} was formerly known as the \textit{EA Sports FIFA} soccer simulation game.}, players are sorted into strength tiers, and probabilities are only reported at the tier level. The suspicion is that players of different strengths are not equally likely within the tiers, but rather, the distribution is skewed towards the lower bound. In \textit{Counter-Strike}, the game displays only the rarity category of possible ``loot'' without disclosing the precise probabilities of obtaining specific items within each tier.\footnote{Due to Chinese regulation, some approximate odds for these rarity tiers are published in that market, but they are not made salient or disclosed within the game at all in regions where such disclosure is not required. See, for instance: \url{https://csgoskins.gg/blog/csgo-case-odds-the-official-numbers-published-by-valve}, accessed on October 22, 2025.} Similar to \emph{EA Sports FC}, within each tier, the distribution remains entirely opaque. Despite this, some of the items obtained through these cases are traded for more than \$1,000,000 on secondary markets.\footnote{See for instance: \url{https://dmarket.com/blog/most-expensive-csgo-skins/}, accessed on October 22, 2025.} FIFA recently started to sell tickets for their soccer tournaments with a so-called Right to Buy (RTB) mechanism. A RTB allows the buyer to purchase a ticket for a FIFA World Cup match. The RTB price only covers the right to buy; it does not cover the ticket itself. Crucially, buyers must complete the purchase of the actual ticket within a designated time window. This time window is announced at short notice. Once it has closed, the RTB expires. However, as there is no information on either the number of RTBs issued per ticket or the length of the time window, the odds of securing a ticket remain uncertain.

\begin{figure}[!ht]
    \centering
    \caption{Clan chat with selective feedback in \textit{Clash Royale}} 
    \includegraphics[width=0.35\textwidth]{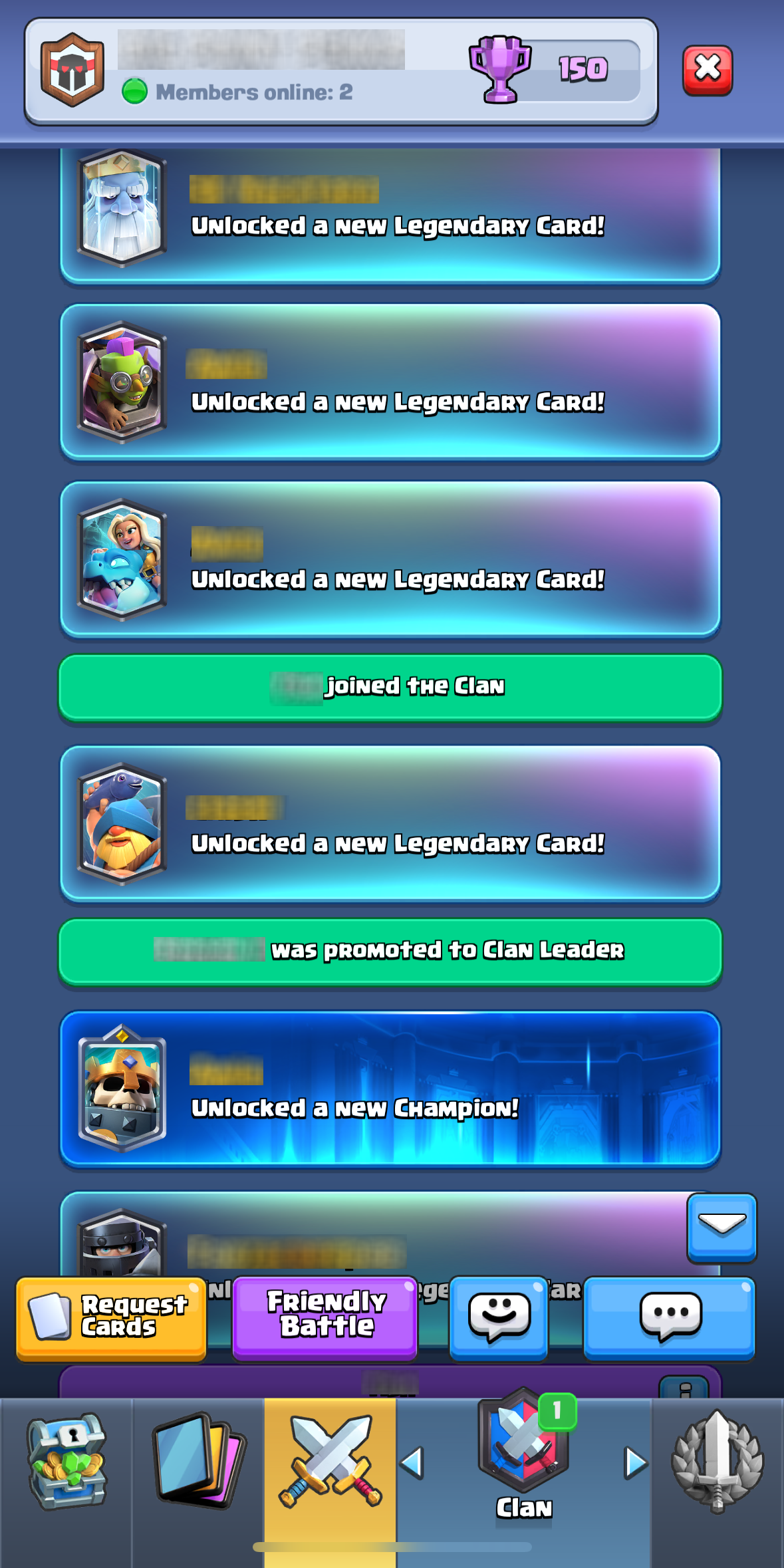}
    \label{fig:Clash Chat}
    \begin{flushleft} 
        \footnotesize
        \setlength{\parskip}{0em} 
        Note: In \textit{Clash Royale} gamers get notified, whenever one of their clan members unlocks a high value item. Names of gamers have been blurred.
    \end{flushleft}
\end{figure}

The second mechanism is selective feedback: rare and valuable outcomes are made highly visible through in-game announcements, social comparison during gameplay, and promotional content. Players often observe only top outcomes won by others, while common or low-value results remain hidden. Content creators reinforce this skew by livestreaming loot box openings and publishing edited highlight reels that feature only the most impressive wins. There are even allegations that sponsored streamers receive preferential odds during promotional events.\footnote{See, for example, \url{https://gamerant.com/ftc-loot-boxes-better-odds-sponsored-streamers/}, accessed on July 20, 2025.} These features enhance engagement but also create belief distortions, thereby increasing the willingness to pay for those digital items and possibly resulting in overspending. One example for selective feedback is the mobile game \textit{Clash Royale}. In the clan chat, gamers get notified whenever one of their clan members receives a high value item for the first time (see Figure~\ref{fig:Clash Chat}). Information about low value items is not shared. 

Our paper investigates spending in consumer lotteries when deceptive features emerge endogenously through seller decisions. Specifically, we analyze how sellers frame their decisions when offering probabilistic products to buyers. The novelty lies in the interaction between sellers and buyers, and in the fact that participants acting as sellers design the lotteries themselves. While the lotteries are generally presented in a neutral manner, without any appealing features that buyers might enjoy, sellers have the option to incorporate deceptive elements commonly found in real-world digital markets. Our lottery market is designed to provide the neutral and sober environment necessary to quantify overspending. Although our experimental design is motivated by practices surrounding loot boxes in video game markets, the mechanisms we study apply more broadly to any environment in which sellers use framing to influence beliefs about probabilistic products.

In an online experiment, our seller subjects have the option of employing two deceptive features. In treatment variations, they may choose to censor the odds of the lotteries and may provide buyers with selective feedback about the success of past ones. We analyze the seller choices and how they affect buyers' willingness to pay for lotteries. We also elicit buyer first-order beliefs and seller second-order beliefs on the probability of winning the highest payout.

Our results from 802 participants show that sellers overwhelmingly choose to present lotteries using deceptive features: across treatments, such presentations are selected in over 80\% of cases. In the treatment that allows both selective sampling and censoring of information, the fully deceptive option is the modal choice. These presentation strategies significantly affect the prices sellers charge and the buyers' willingness to pay. Offer prices are substantially higher when sellers use selective samples, and buyers are willing to pay more for deceptively presented lotteries---up to six times the expected value---compared to transparent ones. This increase in willingness to pay is fully mediated by inflated first-order beliefs about the chance of winning. Sellers also anticipate this effect and adjust their offer prices in line with their second-order beliefs about buyers' perceptions. Together, these findings demonstrate that strategic presentation formats systematically shift consumer beliefs, leading sellers to set higher prices and buyers to assign greater value to the same underlying product.

Our findings have direct implications for marketing research and firms operating in markets with uncertain product outcomes. The ability to influence consumer beliefs through how a product is presented, not by altering the product itself but by changing its perception, offers a low-cost and scalable way to increase perceived value and pricing power. In practice, this creates strong incentives for firms to adopt designs that selectively disclose information, highlight rare successes, or otherwise shift attention toward favorable outcomes. Such strategies may be especially attractive in digital environments, where feedback and framing can be continuously optimized and even personalized. From a managerial perspective, these tools offer a means of extracting additional consumer surplus without changing the underlying odds. This makes them a potentially powerful lever for revenue generation in markets built around probabilistic sales.

These same mechanisms, however, raise concerns from a policy perspective. Our findings contribute to a broader policy discussion on the regulation of consumer products sold through probabilistic mechanisms. Across markets, from digital loot boxes to physical blind-box toys, concerns have grown over designs that obscure probabilities and exploit belief distortions to drive overspending. Rather than supporting outright bans, our results point to transparency requirements as a useful policy tool. Such measures have already been implemented in China, where game publishers must disclose precise success probabilities for loot boxes,\footnote{See \citet{xiao2024gaming}; and \url{https://dmarket.com/blog/most-expensive-csgo-skins/}, accessed July 20, 2025.} and where new national regulations for blind-box toys prohibit sales to young children and require clear disclosure of item odds and return policies.\footnote{See Reuters, China issues rules on mystery boxes, regulates sales to children, June 19, 2023: \url{https://www.reuters.com/article/business/china-issues-rules-on-mystery-boxes-regulates-sales-to-children-idUSL1N3870NF/}; and People’s Daily, June 20, 2025: \url{https://paper.people.com.cn/rmrb/pc/attachement/202506/20/1ba98a62-1b28-48e6-8b0f-c1247ed48d8f.pdf}.} These examples reflect a broader regulatory trend that shifts attention from the use of randomness itself to the strategic ways in which it is framed and disclosed.

The remainder of the paper is organized as follows: Section~\ref{sec:Related literature} discusses the related literature. In Section~\ref{sec:Experimental design}, the experimental design is outlined. Afterwards, in Section~\ref{sec:Hypotheses}, our hypotheses are stated and explained. In Section~\ref{sec:Results} we present the results from our experiment, and in Section~\ref{sec:Conclusion} we conclude.

%% file: Related_Literature.tex
\section{Related literature}
\label{sec:Related literature}

Our study connects to several strands of literature. First, we contribute to the growing literature on gaming, in particular the economics and psychology of loot boxes. Numerous studies have established a positive correlation between self-reported measures of gambling and overspending on loot boxes \citep{drummond2018video, brooks2019associations, zendle2019loot, drummond2020loot}. Theoretical work on loot boxes \citet{chen2020loot, 
hu2024pricing, 
miao2024designing} investigates 
optimal loot box design and pricing under the assumption that buyers maximize expected utility. Empirical research has shown that video game spending, especially on loot boxes, is predictive of excessive credit card use \citep{gong2024debtors}, and that high-spending players systematically differ in preferences from regular players \citep{amano2024makes}. Closest related to us, \citet{cordes2024drives} study the effect of deceptive features in a setting where deceptive features are exogenously imposed by the experimenter. In contrast, our design allows deceptive features to emerge endogenously through sellers’ strategic choices. This difference is crucial: it enables us, first, to study how buyers respond to lottery features when they are aware that these features are chosen by sellers with monetary interests (rather than by experimenters, whose incentives for deception are less apparent), and second, to examine the extent to which sellers actively employ deception when designing lotteries.

Second, our findings speak to the literature on ambiguity attitudes. While ambiguity aversion has long been considered a universal trait, recent work suggests that individuals may be ambiguity-seeking under certain conditions. For instance, \citet{kocher2018ambiguity} and \citet{chandrasekher2022dual} provide evidence that individuals are ambiguity-seeking when winning probabilities are small. Similarly, \citet{golman2021information} argue that people may prefer ambiguity when information gaps are pleasurable to think about, especially when the outcomes are emotionally salient or high in valence. Our results are also in line with source-dependent ambiguity preferences: \citet{baillon2025source} develop a model in which ambiguity attitudes depend on the perceived source of uncertainty. Applying this idea to our context suggests that our experimental design may underestimate real-world ambiguity-seeking behavior, as video game players might perceive ambiguity arising from game mechanics as more enjoyable or exciting than ambiguity in more neutral domains. Consistent with this, \citet{tversky1995weighing} show that basketball fans exhibit ambiguity aversion in Ellsberg-type urn tasks but are ambiguity-seeking when betting on basketball games.

Ambiguous information may also arise from selective disclosure, from which people then draw naive inferences \citep[e.g.,][]{koehler2009selection,deversi2021spin,jin2021nonews,jin2022complex,esponda2018endogenous,barron2024selection,enke2020wysiati}. Our contribution is to show that such selective disclosure can emerge endogenously in a loot-box context.

A related strand of literature in marketing examines how firms strategically design price presentations to induce consumer overspending. For example, \citet{blake2021price} analyze a large-scale field experiment and show that drip-pricing---where mandatory fees are disclosed only late in the purchase process---leads consumers to spend more overall. Evidence that such price complexity can endogenously arise in buyer-seller interactions is provided by \citet{kalayci2015price}, who demonstrate in a laboratory experiment that sellers exploit tariff complexity to confuse buyers and sustain higher prices. These findings belong to the broader literature on obfuscation and shrouded attributes, which highlights how sellers profit from deliberately increasing the cognitive load of buyers \citep[e.g.,][]{ellison2009search,brown2010shrouded}. Our study shares with this literature the focus on deception as an endogenous seller strategy, but differs in an important dimension: in our setting, it is not the price that is shrouded, but the product itself.

Finally, we also relate to the marketing literature on mysterious consumption \citep{buechel2023mysterious}, which investigates whether consumers prefer products whose exact nature is unknown at the time of purchase. It is shown that such preferences hold across a wide range of products. \citet{yin2025probabilistic} theoretically study the economic value of customized mystery box products and explore the interrelationship between probabilistic selling and customization. In contrast, we demonstrate \emph{how} product uncertainty can be framed in a way that leads buyers to overspend.

%% file: Experimental_Design.tex
\section{Experimental design and procedures}
\label{sec:Experimental design}

\paragraph{Basic structure.} In an online experiment, seller and buyer subjects interact on a market for lotteries that resemble loot boxes or other probabilistic products increasingly offered by firms in real-world markets. Sellers offer the lotteries to buyers, possibly using deceptive features. We elicit the sellers' willingness to sell (WTS) and the buyers' willingness to pay (WTP). If the WTP matches or exceeds the WTS, the lottery is sold. 
Each lottery can be written as $(0, 0.99-q; 10, q; x, 0.01)$, where $0$, $10$, and $x$ denote the possible outcomes and $0.99-q$, $q$, and $0.01$ the respective probabilities. In each of five rounds, we independently draw (without replacement) an outcome \( x \in \{100, 120, 140, 160, 180\} \) and a probability \( q \in \{0.09, 0.19, 0.29, 0.39, 0.49\} \), so that 25 different outcome-probability combinations are possible. We also call $x$ the \emph{jackpot} in the following. Each seller/buyer can sell/buy five of these 25 lotteries.

\begin{figure}[!ht]
    \centering
    \caption{A transparent lottery design} 
    \includegraphics[width=0.8\textwidth]{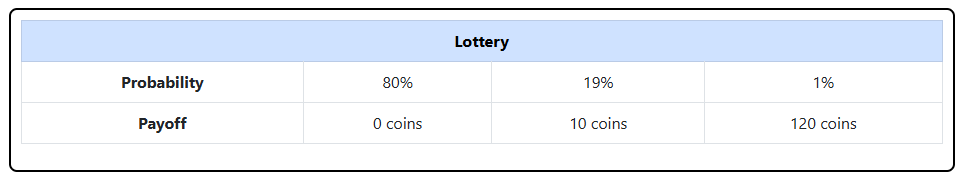}
    \label{fig:transparent_lottery}
    \begin{flushleft} 
        \footnotesize
        \setlength{\parskip}{0em} 
        Note: This lottery design is available in all treatments. In the \textit{No-Choice} treatment, it is the only option.  
    \end{flushleft}
\end{figure}

\paragraph{Treatments.} 
In addition to the baseline variant, which we label \textit{No-Choice}, there are three treatments. We call these \textit{Choice-Sample}, \textit{Choice-Censoring}, and \textit{Choice-Both}. In the three treatments, sellers have multiple options to present the lottery.

In \textit{No-Choice},  sellers cannot decide about the presentation form of the lottery. Buyers are shown all payoffs and odds of the lottery. No additional information on the lottery is provided. Thus, subjects see only the plain lottery. An example of the offered lotteries is displayed in Figure~\ref{fig:transparent_lottery}. It reflects a scenario in which a firm offers a probabilistic good without using any deceptive features.

\begin{figure}[!ht]
\begin{center}
    \caption{A lottery design with a selective sample} 
    \includegraphics[width=0.8\textwidth]{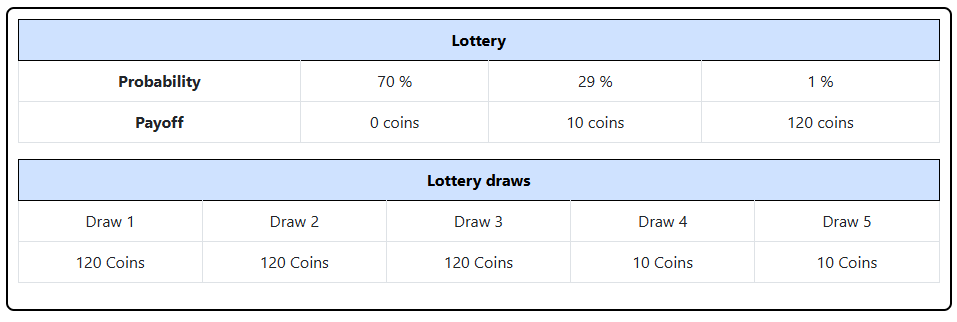}
    \label{fig:lottery_sample}
        \footnotesize
        \setlength{\parskip}{0em} 
\end{center}        
 \footnotesize   
Note: In treatment \textit{Choice-Sample}, participants have two options for the lottery design. They can either choose the selective sample as shown here, or the transparent design from Figure~\ref{fig:transparent_lottery}. The order of both options is randomized. The selective sample is also available in treatment \textit{Choice-Both}. 
\end{figure}

In the treatment \textit{Choice-Sample}, seller subjects can choose between two different presentations.  They can either present the plain lottery from \textit{No-Choice}, or choose to add the five highest payoffs out of 400 random draws to the presentation of the lottery. Since the probability of the highest payout \(x\) is 1\%, these draws may, for example, show \(x\) four times and the medium payoff 10 once. Both sellers and buyers are fully informed of how the draws were generated and are not deceived in any way. The draws are generated before the sellers decide about the presentation format. Figure~\ref{fig:lottery_sample} shows the presentation form with a sample. The order of the presentations is randomized in all treatments. The \textit{Choice-Sample} treatment is motivated by the observation that real-world sellers often highlight rare, desirable outcomes to influence buyer beliefs, even when the underlying probabilities are unchanged.

\begin{figure}[!ht]
\begin{center}
    \caption{A censored lottery design } 
    \includegraphics[width=0.8\textwidth]{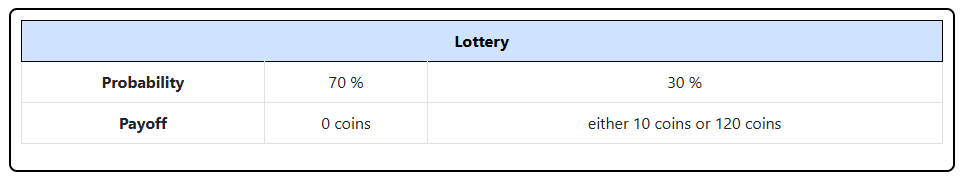}
    \label{fig:lottery_censored}
\end{center}
\footnotesize
\setlength{\parskip}{0em} 
Note: In treatment \textit{Choice-Censoring}, participants have two options for the lottery design. They can either choose the censored lottery presentation as shown here, or the transparent design from Figure~\ref{fig:transparent_lottery}. The order of both options is randomized. The censored lottery presentation is also available in treatment \textit{Choice-Both}. 
\end{figure}

In the treatment \textit{Choice-Censoring}, sellers can again choose between the plain lottery from \textit{No-Choice} and a new option where the odds of the jackpot are censored. In this format, the probabilities of the non-zero payoffs are combined. The buyer will see that there is a probability of \(q+0.01\)  of winning either the jackpot \(x\) or the medium payoff of 10. Figure~\ref{fig:lottery_censored} shows an example lottery from this treatment. This mirrors how firms often hide exact odds to possibly make high rewards seem more likely.

\begin{figure}[!ht]
\begin{center}
    \caption{Lottery design in the \textit{Choice-Both} treatment} 
    \includegraphics[width=0.8\textwidth]{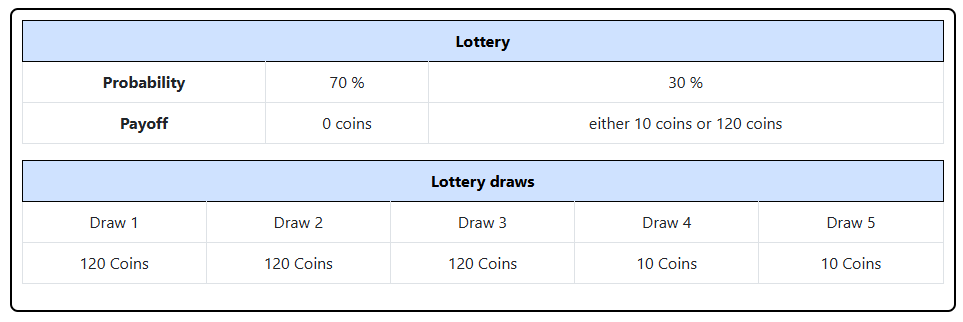}
    \label{fig:lottery_sample+censored}
\end{center}
   \footnotesize
   \setlength{\parskip}{0em} 
        Note: In treatment \textit{Choice-Both}, participants have four options. They have all the options from the other treatments, and additionally, the option to combine both key features of loot boxes. The order of the four different presentations is randomized.
\end{figure}

The last treatment is labeled \textit{Choice-Both}. In this treatment, each seller can choose between four different lottery presentations, presented in random order. The plain lottery from the treatment \textit{No-Choice}, the presentation with an additional sample from the treatment \textit{Choice-Sample}, the presentation with censored odds from the treatment \textit{Choice-Censoring}, and a presentation that combines the two key features of loot boxes and similar lottery-like features. In this fourth presentation, the odds of the lottery are censored as in the treatment \textit{Choice-Censoring} and an additional sample is shown as in the treatment \textit{Choice-Sample}. An example of this presentation form can be seen in Figure~\ref{fig:lottery_sample+censored}. This reflects how firms often combine both deceptive presentations, and which together often represent the status quo in markets for probabilistic goods.

\paragraph{Procedures. } 

The sequence of decisions made by the sellers and buyers is as follows: First, the sellers decide how to present the lottery, and we ask the sellers to briefly justify their decision in a short written comment.\footnote{We provide an analysis of this justification in Table \ref{tab:justification_categories}.} Then, they enter their WTS. We also ask them for their second-order belief of how frequently the buyers believe they would win $x$, and for some sociodemographic data. After seeing the presentation of the lottery, the buyers enter their WTP and their first-order belief of how often they believe they would win $x$. We also ask buyers for sociodemographic data.

One seller and one buyer are matched with each other for the entire experiment, and there are five rounds in which lotteries can be sold for a seller-buyer pair. The sellers decide on five different lotteries in a row, receiving an independent draw without replacement from the above set of possible lotteries. They do so without feedback or interaction with the buyers. When the buyers decide on their WTP for each of the five lotteries, they receive feedback on whether they bought the lottery in each round. The feedback includes their stated WTP, the price of the lottery, and a potential bonus payment in this round. 

We incentivize the seller-buyer interaction: One out of every seven seller-buyer pairs is randomly selected to be eligible for a bonus payment. For these seller-buyer pairs, one round is randomly selected to be bonus relevant. If the lottery is sold (that is the WTP is at least as high as the offer price), the potential bonus payment for the seller is the offer price. The potential bonus payment for the buyer is determined by a random draw of the lottery. If the offer price is higher than the WTP, the seller does not receive a bonus payment, and the buyer’s potential bonus payment is the offer price. We use the BDM mechanism \citep{becker1964measuring} to incentivize subjects to report their true willingness to pay.

We use a between-subjects design for the sellers, meaning each seller is always allocated to exactly one treatment. The buyers receive all five lotteries from a single seller, so in a way, they are allocated to one treatment, too. However, the  presentation chosen by the seller may differ. Theoretically, a buyer matched with a seller from the \textit{Sample\_Censoring} treatment could see the lotteries presented in four different formats. Therefore, we analyze sellers' data based on the treatment, but buyers' decisions based on the format of the lottery presented to them rather than the treatment. 

\paragraph{Beliefs and demographics. } 
Before eliciting sellers' WTS, we collect second-order beliefs by asking sellers how often they think that a buyer would expect to win the highest payoff, $x$, out of 100 times. Buyers are shown the lottery in the frame chosen by the sellers. We ask them for their WTP and their first-order beliefs how often they think they expect to win the highest payoff, $x$. Recent evidence suggests that standard incentivization techniques systematically distort reported beliefs \citep{danz2022belief}. Therefore, we do not incentivize beliefs.

After collecting our decision data, we survey basic demographics, measures of loot box spending, measures of self-control \citep{tangney2004high}, and measures of gambling behavior \citep{ferris2001canadian}. This helps to test the external validity of our experiment, as we can compare survey responses with behavior in our task. For instance, people with lower self-control might be more likely to overspend. Participants with gambling habits could be more attracted to lottery-like products. Moreover, those who spend more on loot boxes, which are a major form of such products, may also be more willing to pay for the lotteries in our experiment.

\paragraph{Implementation. } 
The design was pre-registered at OSF Registries\footnote{See  \url{https://doi.org/10.17605/OSF.IO/G7P3X}}. We collected data from 802 subjects (401 sellers and 401 buyers) located in the United Kingdom (UK) via \textit{Prolific} in May 2025. To prevent access by automated bots, we implemented Google reCAPTCHA, a tool that distinguishes between human and automated responses, on the welcome page of the experiment \citep{rilla2025recognising}. We screened out inattentive participants with an attention check at the beginning of the experiment. To verify that our instructions (which can be found in Appendix B) were understood, we asked multiple comprehension questions after the instructions. Only subjects that answered all questions correctly were allowed to proceed with the experiment. Approximately 75\% of the subjects correctly answered the comprehension questions on their first attempt, while around 11\% were filtered out for answering the questions incorrectly more than once. The base fee for completing the experiment was \pounds2. In addition, 1 out of 7 participants received a bonus payment. The average bonus payment for buyers was \pounds2 for buyers and \pounds0.4 for sellers. The experiment took, on average, 14 minutes to complete.

%% file: Hypotheses.tex
\section{Hypotheses}
\label{sec:Hypotheses}
First, we want to analyze whether the seller subjects actually use the deceptive features when given the option. Since firms in real markets are suspected of adopting such features to increase revenues, we expect our sellers to do the same. Presumably, sellers anticipate that they will be able to sell the lottery for higher prices by using the deceptive features. We hypothesize: 

\bigskip

\noindent \textbf{Hypothesis 1:} \textit{Sellers use the deceptive features when presenting the lottery to the buyer. In the treatment Choice-Both, the presentation that combines both deceptive features is the most frequently chosen one.}

\bigskip

We then focus on the offer prices. In real markets, deceptive features may allow firms to sustain higher prices for probabilistic products, since consumers overestimate their chances of receiving rare items. Similarly, in our experiment, we expect sellers who employ the deceptive features to charge higher prices than those in the treatment \textit{No-Choice}. The prices in the treatment \textit{Choice-Both} should be the highest because the sellers in this treatment can combine both deceptive features. Thus, we hypothesize:

\bigskip

\noindent \textbf{Hypothesis 2:} \textit{The offer price is higher in the treatments Choice-Sample and Choice-Censoring than in the baseline treatment No-Choice. The offer price is higher in Choice-Both than in the treatments Choice-Sample and Choice-Censoring.}

\bigskip

Then we turn to the willingness to pay on the buyer side. We expect the reported willingness to pay when offered a lottery with a deceptive feature to be higher than when offered the lottery without a deceptive feature. We anticipate the willingness to pay for the lotteries with two deceptive features to be higher than the willingness to pay for the lotteries with only one deceptive feature. 
We hypothesize:

\bigskip

\noindent \textbf{Hypothesis 3:} \textit{The WTP for lotteries with one deceptive feature is higher than the WTP for the transparent lottery. The WTP for lotteries with both deceptive features is higher than for lotteries employing only one deceptive feature.}

\bigskip

We anticipate that second-order beliefs are the main driver of increased offer prices by the seller subjects. We hypothesize that sellers expect the buyer to overestimate the likelihood of a jackpot when they are offered a lottery with a deceptive feature.

\bigskip

\noindent \textbf{Hypothesis 4:} \textit{Seller subjects hold the higher second-order beliefs, the more deceptive features are used. The higher the seller's second-order beliefs are, the higher their offer price.}

%% file: Results.tex
\section{Results}
\label{sec:Results}
Our sample consists of 401 sellers and 401 buyers, each of whom made five lottery choices, yielding a total of 2,005 buyer and seller observations. Since each buyer is matched with exactly one seller, we cluster standard errors at the seller level in the following.

\subsection{Seller Results}

\begin{figure}[!htbp]
    \centering
    \caption{Seller choices regarding the presentation of the lotteries} 
    \vspace{0.5em}
    \includegraphics[width=0.6\textwidth]{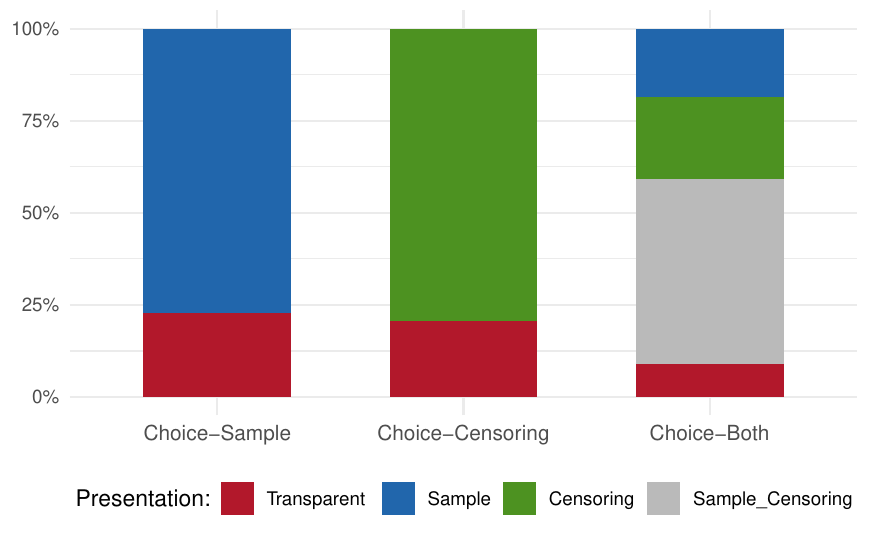}
    \label{fig:choices}
    \begin{flushleft} 
        \footnotesize
        \setlength{\parskip}{0em} 
        Note: In \textit{Sample} and \textit{Censoring} seller subjects had two options whereas in \textit{Sample\_Censoring} seller subjects had four options.
    \end{flushleft}
\end{figure}

First, we analyze the seller's decisions regarding the presentation of the lotteries.  Figure~\ref{fig:choices} shows the distribution of choices. We observe that transparent presentation is chosen rarely, in fewer than 25\% of cases. In \textit{Choice-Both}, the presentation that combines both a selective sample and censoring of the odds is the modal presentation. 

For statistical analysis, we create a dummy variable that is equal to one if the seller chose a lottery presentation with at least one deceptive feature. We then run separate OLS regressions for all three treatments with this dummy as the dependent variable (see Table~\ref{tab:deceptive feature regression}). The columns (1) and (2) of Table~\ref{tab:deceptive feature regression} show that in the \textit{Choice-Sample} and \textit{Choice-Both} treatments, the option with the deceptive feature is chosen around 80\% of the time. Wald tests show that this is significantly more often than 50\%, that is, more often than the plain transparent lottery ($p<0.001$), see the bottom of columns (1) to (3). In the treatment \textit{Choice-Both} a presentation with at least one deceptive feature is chosen 91\% of the time (see column (3) of Table~\ref{tab:deceptive feature regression}), also more often than the plain lottery ($p<0.001$). 
Additionally, we perform a multinomial logit with clustered standard errors in column (4) of Table~\ref{tab:deceptive feature regression}. The reference category is the presentation form with two deceptive features. We see that all other presentation forms are chosen significantly less often ($p<0.01$). This is support for our first hypothesis, summarized in Result 1. 

\begin{table}[!htbp] \centering 
  \caption{Regression results - deceptive features} 
  \label{tab:deceptive feature regression}
  \small

\resizebox{\textwidth}{!}{
\begin{tabular}{@{\extracolsep{5pt}}lcccc} 
\\[-1.8ex]\hline 
\hline \\[-1.8ex] 
\\[-1.8ex] & \multicolumn{3}{c}{\textit{OLS}} & \textit{Mlogit} \\
\cline{2-4} \cline{5-5}
\\[-1.8ex] & Choice\text{-}Sample & Choice\text{-}Censoring & Choice\text{-}Both & Choice\text{-}Both \\ 
 & (1) & (2) & (3) & (4)\\ 
\hline \\[-1.8ex] 
Constant & 0.771$^{***}$ & 0.794$^{***}$ & 0.910$^{***}$ & \\
 & (0.030) & (0.033) & (0.021) & \\
Censoring &  &  &  & $-$0.816$^{***}$ \\
 &  &  &  & (0.114) \\
Sample &  &  &  & $-$0.993$^{***}$ \\
 &  &  &  & (0.122) \\
Transparent &  &  &  & $-$1.719$^{***}$ \\
 &  &  &  & (0.162) \\
\hline \\[-1.8ex]
\textit{Wald tests constant $=$ 0.5:} & & & & \\ 
$p$-values & $<0.001$ & $<0.001$ & $<0.001$ & \\
\hline
Observations & 480 & 495 & 500 & 500 \\
\hline 
\hline \\[-1.8ex]  
\end{tabular}
}

\vspace{0.5em}
\begin{minipage}{\textwidth}
\footnotesize
\setlength{\parskip}{0em}
\textit{Notes}: Results from OLS regressions for all treatments. The dependent variable is a dummy variable called ``deceptive feature'' in all columns, which is equal to one if the seller chose a lottery presentation with a deceptive feature. In column (3), the deceptive feature can be either the selective sample, the censored odds, or both combined. In column (4), we perform a multinomial logit with ``Sample\_Censoring" as reference category. Standard errors clustered at the seller level are in parentheses. For columns (1) to (3), Wald tests show that lotteries with deceptive features are chosen significantly more than 50\% of the time. \textit{ *~$p<0.1$, **~$p<0.05$, ***~$p<0.01$}
\end{minipage}
\end{table}

\bigskip

\noindent \textbf{Result 1:} \textit{Sellers choose to sell the lotteries significantly more often with deceptive features than without them. In \textit{Choice-Both}, the presentation with both deceptive features is chosen significantly more often than the other three options. }

\bigskip

\begin{figure}[!htbp]
\centering
\caption{Average offer prices and second-order beliefs by treatment}
\vspace{0.5em}
\begin{subfigure}[t]{0.48\textwidth}
    \centering
    \includegraphics[width=\textwidth]{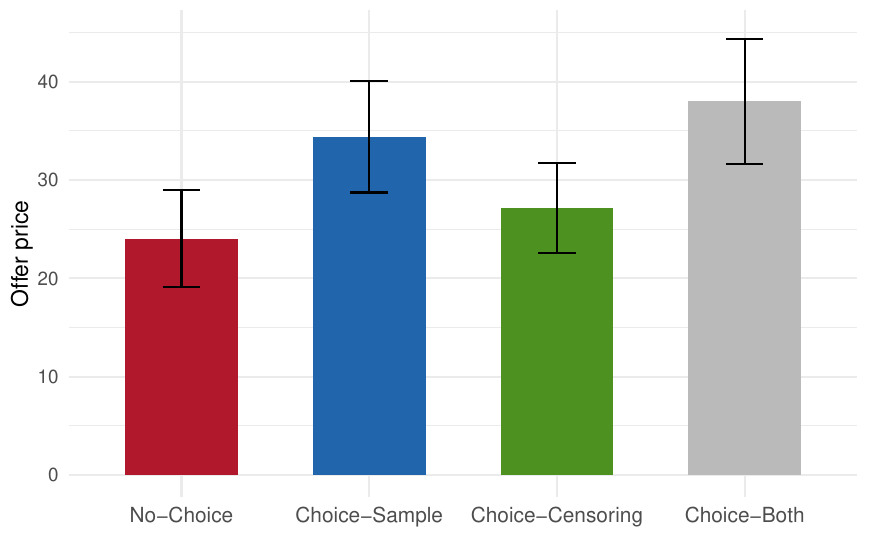}
\end{subfigure}
\hfill
\begin{subfigure}[t]{0.48\textwidth}
    \centering
    \includegraphics[width=\textwidth]{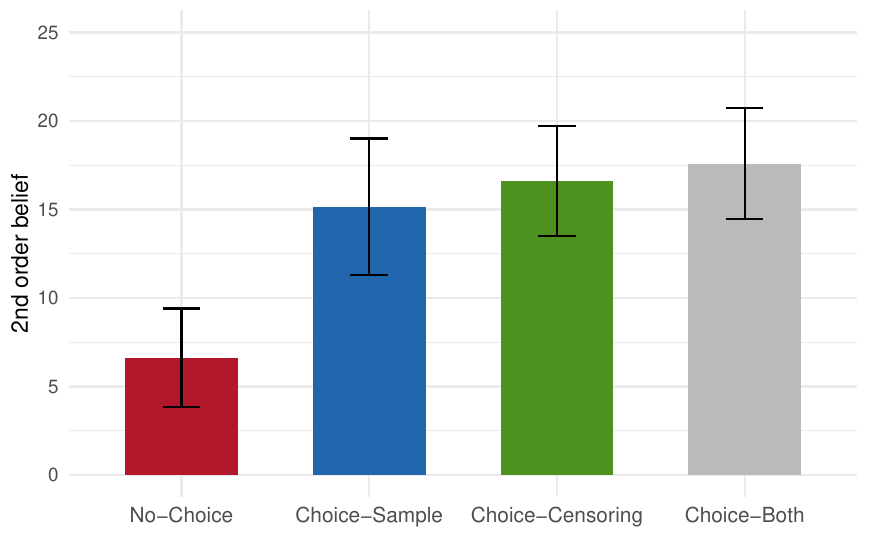}
\end{subfigure}
\label{fig:seller}
\begin{flushleft} 
        \footnotesize
        \setlength{\parskip}{0em} 
        Left panel: mean offer prices in the different treatments. Right panel: mean second-order beliefs in the different treatments. Whiskers are the 95\% CI.  
    \end{flushleft}
\end{figure}

Figure~\ref{fig:seller} displays the mean offer prices (the elicited WTS) posted by sellers in the left panel. We find that seller subjects choose higher offer prices in the two treatments that include the deceptive feature ``selective sample''. Since offer prices are incentivized, this suggests that our seller subjects assume higher WTP in treatments with that deceptive feature. Adding a selective sample increases the offer prices by around 50\%. 

In the treatment \textit{Choice-Censoring}, the mean offer prices are not significantly higher than in the baseline treatment \textit{No-Choice}. If we look at median offer prices (see Figure~\ref{fig:median offer}), we find that in treatments with one deceptive feature, the median offer price doubles compared to the baseline treatment \textit{No-Choice}. This also includes the deceptive feature on censoring the odds. In the treatment \textit{Choice-Both}, the median offer price is three times higher than in the baseline treatment. This is partial support for our second hypothesis: We find higher medians in all treatments with deceptive features. However, there are no significantly higher offer prices in the treatment \textit{Choice-Censoring}.

\bigskip

\noindent \textbf{Result 2:} \textit{Offer prices are significantly higher in treatments with a selective sample.}

\bigskip

The mean second-order belief in all treatments with deceptive features is significantly higher than in the baseline treatment \textit{No-Choice} (see Table~\ref{tab:regression seller}). The right panel of Figure~\ref{fig:seller} shows the mean second-order beliefs. The highest second-order beliefs are reported in the treatment \textit{Choice-Both}. However, the beliefs in the treatment \textit{Choice-Both} are statistically indistinguishable from beliefs in the other treatments with deceptive features. Looking at absolute values, seller subjects anticipate that buyers overestimate the likelihood of a jackpot in the treatment \textit{No-Choice} already. The mean second-order belief in the baseline treatment is around seven times higher than the expected value of winning the jackpot out of 100 times. Thus, the deceptive features increase an already inflated second-order belief even further. 

\begin{table}[!htbp]
\centering
\caption{Regression results -- main specification seller}
\label{tab:regression seller}

\resizebox{\textwidth}{!}{
\begin{tabular}{@{\extracolsep{5pt}}lccccc}
\\[-1.8ex]\hline
\hline \\[-1.8ex]
 & \multicolumn{3}{c}{Offer prices} & \multicolumn{2}{c}{2nd order belief} \\
\cline{2-4} \cline{5-6}
 & (1) & (2) & (3) & (4) & (5) \\
\hline \\[-1.8ex]
Choice-Sample & 10.342$^{***}$ & 9.601$^{**}$ & 4.278 & 8.539$^{***}$ & 8.530$^{***}$ \\ 
 & (3.814) & (3.792) & (3.336) & (2.421) & (2.463) \\ 
Choice-Censoring & 3.106 & 2.398 & $-$3.694 & 9.997$^{***}$ & 9.763$^{***}$ \\ 
 & (3.425) & (3.313) & (3.084) & (2.122) & (2.137) \\ 
Choice-Both & 13.941$^{***}$ & 13.899$^{***}$ & 7.036$^{*}$ & 10.983$^{***}$ & 10.997$^{***}$ \\ 
 & (4.079) & (4.081) & (3.667) & (2.137) & (2.160) \\ 
Belief &  &  & 0.624$^{***}$ &  &  \\ 
 &  &  & (0.072) &  &  \\ 
Constant & 24.043$^{***}$ & 30.434$^{***}$ & 24.229$^{***}$ & 6.613$^{***}$ & 9.943$^{**}$ \\ 
 & (2.502) & (6.055) & (5.400) & (1.421) & (4.345) \\ 
\hline \\[-1.8ex]
\textit{Post-hoc Wald tests:} & & & & & \\ \\ 
Choice-Sample - Choice-Censoring & 7.236$^{*}$ & 7.202$^{**}$ & 7.972$^{**}$ & $-$1.458 & $-$1.233 \\ 
 & (3.709) & (3.546) & (3.240) & (2.515) & (2.501) \\ 
Choice-Both - Choice-Sample & 3.599 & 4.298 & 2.759 & 2.444 & 2.467 \\ 
 & (4.320) & (4.294) & (3.863) & (2.527) & (2.539) \\ 
Choice-Both - Choice-Censoring & 10.835$^{***}$ & 11.501$^{***}$ & 10.730$^{***}$ & 0.986 & 1.234 \\ 
 & (3.981) & (3.861) & (3.557) & (2.243) & (2.223) \\ 
\hline \\[-1.8ex]
Controls & No & Yes & Yes & No & Yes \\ 
2nd order belief & No & No & Yes & No & No \\ 
Lottery FE & No & Yes & Yes & No & Yes \\ 
Observations & 2,005 & 2,005 & 2,005 & 2,005 & 2,005 \\ 
R\textsuperscript{2} & 0.031 & 0.078 & 0.222 & 0.048 & 0.077 \\ 
Adjusted R\textsuperscript{2} & 0.029 & 0.064 & 0.209 & 0.047 & 0.063 \\ 
\hline
\hline \\[-1.8ex]
\end{tabular}
}

\vspace{0.5em}
\begin{minipage}{\textwidth}
\footnotesize
\setlength{\parskip}{0em}
\textit{Notes}: Results from ordinary least squares (OLS) on the treatment dummies. The outcome variable in columns (1), (2), and (3) is the offer price and in (4) and (5) the second-order belief. Columns (1) and (4) do not include control variables. Columns (2), (3), and (5) control for age, gender, and monthly available budget. Standard errors clustered at the seller level in parentheses. The second part of the table shows the results of post-hoc Wald tests. \textit{*~$p<0.1$, **~$p<0.05$, ***~$p<0.01$}
\end{minipage}
\end{table}

\begin{figure}[!htbp]
    \centering
    \caption{Mean offer prices conditional on sellers' choice} 
    \vspace{0.5em}
    \includegraphics[width=0.6\textwidth]{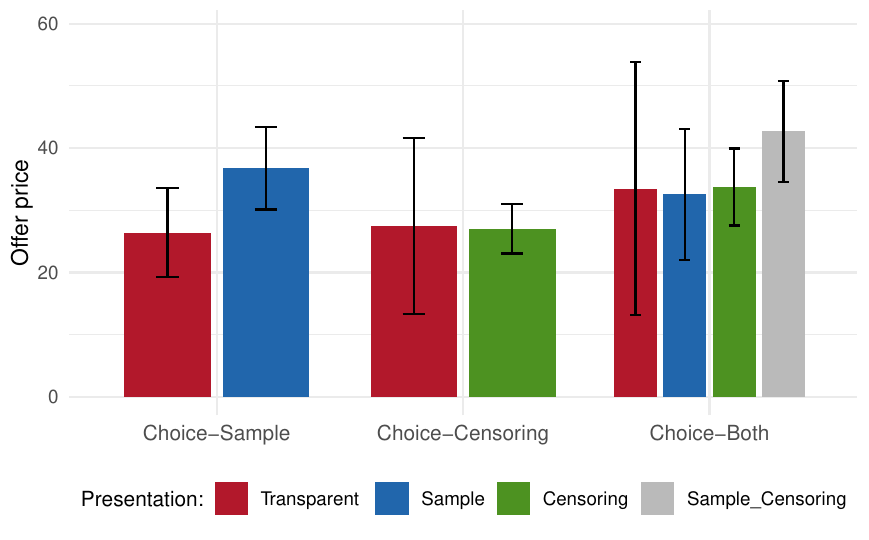}
    \label{fig:offer prices conditional on presentation}
    \begin{flushleft} 
        \footnotesize
        \setlength{\parskip}{0em} 
        Note: Mean offer prices in the different treatments conditional on sellers' choice. In \textit{Choice-Sample} and \textit{Choice-Censoring} seller subjects had two options whereas in \textit{Choice-Both} seller subjects had four options.
    \end{flushleft}
\end{figure}

Next, we analyze offer prices conditional on the sellers' presentation choice. The results are displayed in Figure~\ref{fig:offer prices conditional on presentation}. In the treatment \textit{Choice-Sample}, offer prices are significantly higher when the seller subject chose the presentation with the sample. In the treatment \textit{Choice-Censoring}, seller subjects chose similar prices for both presentation forms. We find the highest prices in the treatment \textit{Choice-Both} when seller subjects chose the presentation form with both deceptive features. Since we only have few observations in some treatment x presentation form combinations (i.e. 45 times Transparent in the treatment \textit{Choice-Both}), confidence intervals are relatively large.

\begin{figure}[!htbp]
    \centering
    \caption{Mean second-order beliefs conditional on sellers' choice} 
    \vspace{0.5em}
    \includegraphics[width=0.6\textwidth]{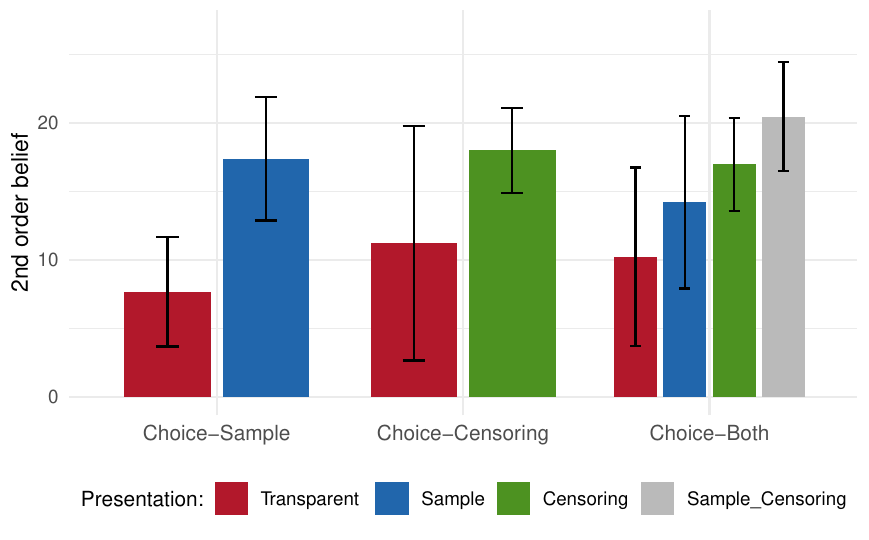}
    \label{fig:second order beliefs conditional on presentation}
    \begin{flushleft} 
        \footnotesize
        \setlength{\parskip}{0em} 
        Note: Mean second-order beliefs in the different treatments conditional on sellers' choice. In \textit{Choice-Sample} and \textit{Choice-Censoring} seller subjects had two options whereas in \textit{Choice-Both} seller subjects had four options.
    \end{flushleft}
\end{figure}

For second-order beliefs, results of the analysis conditional on sellers' choice are displayed in Figure~\ref{fig:second order beliefs conditional on presentation}. In the treatments \textit{Choice-Sample} and \textit{Choice-Censoring}, second-order beliefs are significantly higher when a presentation form with a deceptive feature is chosen. In the treatment \textit{Choice-Both}, the highest second-order beliefs are reported for the presentation form with two deceptive features. The lowest second-order beliefs are reported for the transparent presentation without a deceptive feature. 

\subsection{Buyer Results}

Next, we analyze the decisions of our buyers. Each buyer is matched to one seller for all five rounds. Since the seller subjects in the treatments \textit{Choice-Sample}, \textit{Choice-Censoring}, and \textit{Choice-Both} had multiple options for presenting the lottery in each of the five rounds, buyer subjects in those treatments may have seen multiple presentation forms. 79\% of buyer subjects in the treatment \textit{Choice-Both} see more than one presentation form in the five rounds (see Table~\ref{tab:consistency}). Therefore, we will analyze buyers' decisions based on presentation form, which gives us a within-subjects design for buyers. However, 52.1\% of buyers in the treatment \textit{Choice-Sample}, 70.7\% of buyers in the treatment \textit{Choice-Censoring}, and 100\% of buyers in the treatment \textit{No-Choice} see the same presentation form in all five rounds.  

\begin{figure}[!htbp]
\centering
\caption{Average WTP and first-order beliefs by presentation form}
\vspace{0.5em}
\begin{subfigure}[t]{0.48\textwidth}
    \centering
    \includegraphics[width=\textwidth]{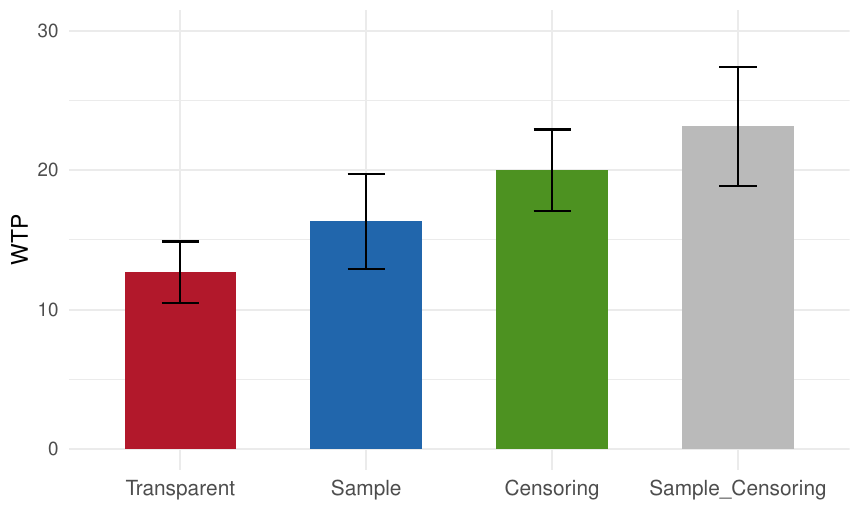}
\end{subfigure}
\hfill
\begin{subfigure}[t]{0.48\textwidth}
    \centering
    \includegraphics[width=\textwidth]{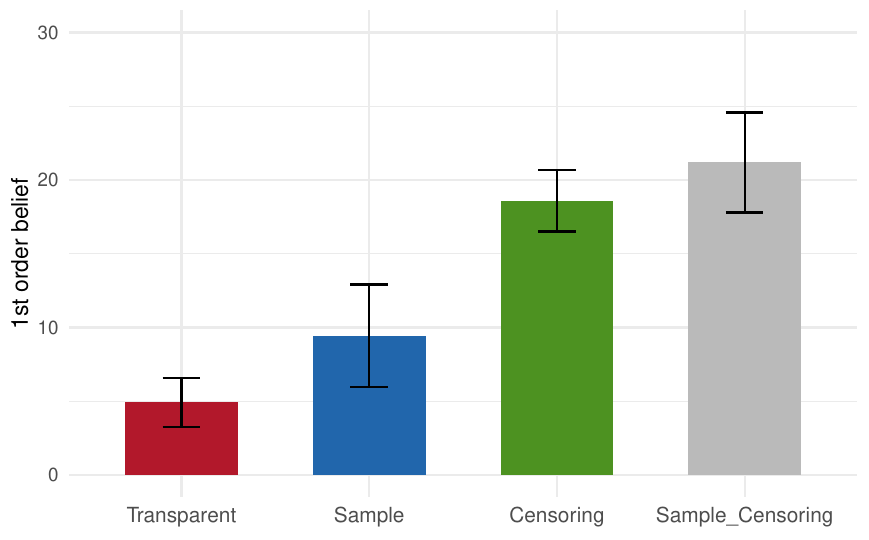}
\end{subfigure}
\label{fig:buyer}
\begin{flushleft} 
        \footnotesize
        \setlength{\parskip}{0em} 
        The left panel shows the mean WTP for the different presentation forms. The right panel shows the mean first-order beliefs for the different presentation forms. Whiskers are the 95\% CI. 
    \end{flushleft}
\end{figure}

The left panel of Figure~\ref{fig:buyer} shows the average WTP by presentation form. The WTP for a lottery with both deceptive features is twice as high compared to the baseline transparent presentation form. Each deceptive feature individually also increases the WTP. However, the censoring of the odds results in a larger increase than adding the highly selective sample. When controlling for age, gender, monthly available budget, and lottery fixed effects, we still find significant effects of the presentation forms (see column (2) of Table~\ref{tab: regression buyer}). When we additionally control for the first-order belief, we no longer find a significant effect of the presentation forms. The first-order belief has a significant effect on the WTP ($p<0.01$). This supports our third hypothesis. Buyers are willing to spend around three times the expected value of the lottery when seeing a transparent lottery. If the lottery is presented with one or two deceptive features, buyers' WTP is four to six times higher than the expected value.

\bigskip

\noindent \textbf{Result 3:} \textit{Buyers have a significantly higher WTP for lotteries with deceptive features.}

\bigskip

Beliefs follow a similar pattern as the WTP. However, the magnitude of the increases is larger. Adding a highly selective sample doubles the mean first-order belief (see right panel of Figure~\ref{fig:buyer}). The first-order belief for a lottery with censored odds or both deceptive features is around 4 times as high as for the baseline lottery. When controlling for age, gender, monthly available budget, and lottery fixed effects, we find that all presentation forms with at least one deceptive feature significantly increase the first-order belief (see column (5) of Table~\ref{tab: regression buyer}). 

\begin{table}[!htbp] \centering
\caption{Regression results -- main specification buyer}
\label{tab: regression buyer}

\resizebox{\textwidth}{!}{
\begin{tabular}{@{\extracolsep{5pt}}lccccc}
\\[-1.8ex]\hline 
\hline \\[-1.8ex]
  & \multicolumn{3}{c}{WTP} & \multicolumn{2}{c}{1st order belief} \\
\cline{2-4} \cline{5-6}
 & (1) & (2) & (3) & (4) & (5)\\  
\hline \\[-1.8ex]  
Censoring & 7.312$^{***}$ & 6.995$^{***}$ & 0.691 & 13.680$^{***}$ & 13.735$^{***}$ \\
 & (1.790) & (1.654) & (1.887) & (1.310) & (1.252) \\
Sample & 3.653$^{*}$ & 3.554$^{*}$ & 1.508 & 4.512$^{**}$ & 4.458$^{**}$ \\
 & (1.990) & (2.019) & (1.874) & (1.960) & (1.951) \\
Sample\_Censoring & 10.454$^{***}$ & 10.628$^{***}$ & 3.002 & 16.270$^{***}$ & 16.618$^{***}$ \\
 & (2.439) & (2.443) & (2.585) & (1.920) & (1.904) \\
Belief &  &  & 0.459$^{***}$ &  &  \\
 &  &  & (0.075) &  &  \\
Constant & 12.682$^{***}$ & 11.975$^{***}$ & 10.007$^{***}$ & 4.916$^{***}$ & 4.288 \\
 & (1.123) & (3.064) & (2.792) & (0.856) & (2.721) \\
\hline \\[-1.8ex]
\textit{Post-hoc Wald tests:} & & & & & \\ \\ 
Sample - Censoring & $-$3.658 & $-$3.441 & 0.817 & $-$9.168$^{***}$ & $-$9.277$^{***}$ \\
 & (2.283) & (2.208) & (2.253) & (2.071) & (2.022) \\
Sample\_Censoring - Sample & 6.801$^{***}$ & 7.074$^{***}$ & 1.493 & 11.758$^{***}$ & 12.160$^{***}$ \\
 & (2.542) & (2.514) & (2.658) & (2.448) & (2.430) \\
Sample\_Censoring - Censoring & 3.142 & 3.634 & 2.310 & 2.590 & 2.883 \\
 & (2.558) & (2.441) & (2.343) & (1.991) & (1.910) \\
\hline \\[-1.8ex]
Controls & No & Yes & Yes & No & Yes \\ 
1st order belief & No & No & Yes & No & No \\ 
Lottery FE & No & Yes & Yes & No & Yes \\ 
Observations & 2,005 & 2,005 & 2,005 & 2,005 & 2,005 \\ 
R$^{2}$ & 0.027 & 0.073 & 0.172 & 0.140 & 0.185 \\ 
Adjusted R$^{2}$ & 0.026 & 0.059 & 0.158 & 0.139 & 0.172 \\ 
\hline 
\hline \\[-1.8ex]
\end{tabular}
}
\vspace{0.5em}
\begin{minipage}{\textwidth}
\footnotesize
\setlength{\parskip}{0em}
\textit{Notes}: Results from ordinary least squares (OLS) on the presentation form dummies. The outcome variable in columns (1), (2), and (3) is the WTP and in (4) and (5) the first-order belief. Columns (1) and (4) do not include control variables. Columns (2), (3), and (5) control for age, gender, and monthly available budget. Standard errors clustered at the buyer level in parentheses. The second part of the table shows the results of post-hoc Wald tests. \textit{ *~$p<0.1$, **~$p<0.05$, ***~$p<0.01$}
\end{minipage}
\end{table}

Sellers overestimate the buyers' beliefs for the baseline lottery and the presentation form with sample (see Figure~\ref{fig:belief seller presentationform}). For the presentation with both deceptive features and the presentation with censored odds, sellers' slightly underestimate the buyers' beliefs. However, seller subjects still correctly anticipate that the first-order belief is increased due to deceptive features. There is also a significant effect of the sellers' second-order beliefs on the offer prices (see column (3) Table~\ref{tab:regression seller}). Thus, our results also confirm our fourth hypothesis.

\bigskip

\noindent \textbf{Result 4:} \textit{First- and second-order beliefs significantly increase WTP and offer price, respectively.}

\subsection{Mechanism: Overoptimistic beliefs drive high WTPs}

Offer prices are, on average, around five to nine times higher than the expected value. Even when considering median values, offer prices remain around two to seven times higher. This is not a laboratory artefact, though, as probabilistic product offers are frequently sold at prices that exceed their expected value. Consider, for instance, EA's FC (formerly FIFA) franchise, where the price of the ``Premium Gold Pack'' is about three times higher than the expected value of the items it contains in in-game currency.\footnote{The ``Premium Gold Pack'' is offered for 7,500 coins or 150 FC Points (approximately \$1.25). The average value based on 15,500 packs is 2,701 coins (see \url{https://futmind.com/pack/301/premium-gold-pack}, accessed on July 10th, 2025).} When converted to real-world money, the discrepancy becomes even more pronounced, as the price is between three and 23 times higher than the expected value.\footnote{On the third-party market, 2.5 million coins were worth around \$400 on the release day of EA FC 25 and less than \$100 by February 2025 (see \url{https://www.playerauctions.com/market-price-tracker/ea-fc/}, accessed on July 10th, 2025). Generating 2.5 million coins would require the purchase of approximately 925 ``Premium Gold Packs'', costing around \$1,150.} Our results provide insights into the factors driving this large wedge between offer prices and expected values.

\begin{table}[!htbp] \centering 
  \caption{Difference between WTP and perceived expected value} 
  \label{tab:perceived_value} 
  \resizebox{\textwidth}{!}{
\begin{tabular}{@{\extracolsep{5pt}} lccccc} 
\\[-1.8ex]\hline 
\hline \\[-1.8ex] 
& \multicolumn{5}{c}{Presentation form} \\
\cline{2-6}
\\[-1.8ex]
 & All & Transparent & Sample & Censoring & Sample\_Censoring \\ 
 & (1) & (2) & (3) & (4) & (5)\\
\hline \\[-1.8ex] 
Mean WTP & $16.586$ & $12.682$ & $16.336$ & $19.994$ & $23.136$ \\ 
Mean perceived EV & $17.601$ & $9.139$ & $14.861$ & $27.455$ & $30.843$ \\ 
Difference & $-1.015$ & $3.544$ & $1.474$ & $-7.461$ & $-7.707$ \\ 
\hline
\hline \\[-1.8ex] 
\end{tabular} 
}
\vspace{0.5em}
\begin{minipage}{\textwidth}
\footnotesize
\setlength{\parskip}{0em}
\textit{Notes}: Mean WTP and mean perceived expected value (EV) for all presentation forms combined (column (1)) and each presentation form separately (columns (2) to (5)).
\end{minipage}
\end{table}

We can calculate the perceived expected value of the lottery by eliciting buyers' first-order beliefs, which yields the following expression:  
\[
\text{perceived EV} = 10 \cdot (q - \text{belief}) + x \cdot \text{belief}.
\]  
We find that the average difference between the perceived expected value and WTP is small. Across all presentation formats, the difference between mean WTP and mean perceived expected value is only 1.015 coins (see Table~\ref{tab:perceived_value}, column (1)). At first glance, this suggests that subjects do not consciously overpay, as WTP is close to the perceived expected value. However, the deceptive features shift beliefs upward, thereby inducing overspending. This finding has important policy implications: it indicates that overspending cannot be explained simply by preferences for gambling but rather results from overoptimistic beliefs induced by the framing of the lottery, which provides a rationale for regulatory intervention.

\subsection{External validity} To assess the external validity of our results, we examine the relationship between our experimental measures and real-world loot box overspending. In the survey accompanying our experiment, we first asked subjects whether they know what loot boxes are. Among buyer subjects, 71\% reported familiarity. For these subjects, we elicited both their loot box spending and whether they had ever spent more on loot boxes than they had planned. Overall, 20\% reported having overspent on loot boxes. We define \emph{overspenders} as those who indicated spending more than planned.  

Ninety of our buyer subjects reported positive loot box spending in the past 12 months. On average, these buyers spent more than \pounds 45 per month, although this mean is driven by a small group of individuals. The median spending is substantially lower, at \pounds 20 per month.  

Table~\ref{tab:overspending} reports results from OLS regressions. Column (1) shows that a higher WTP in our experiment significantly predicts the likelihood of real-world overspending, whereas beliefs do not. In addition, measures of gambling behavior significantly increase the probability of overspending on loot boxes. Table~\ref{tab:loot box spending} further shows that both WTP and gambling behavior are strong predictors of the level of loot box spending. These correlations suggest that our findings are indeed informative about real-world behavior.

\begin{table}[!htbp] \centering 
  \caption{Predictors for real world overspending - regression results} 
  \label{tab:overspending} 
  {\small
\begin{tabular}{@{\extracolsep{5pt}}lcccc} 
\\[-1.8ex]\hline 
\hline \\[-1.8ex] 
 & \multicolumn{4}{c}{Overspending (Dummy)} \\ 
\cline{2-5} 
\\[-1.8ex] & (1) & (2) & (3) & (4)\\ 
\hline \\[-1.8ex] 
 Constant & 0.152$^{***}$ & 0.177$^{***}$ & 0.152$^{***}$ & 0.118$^{***}$ \\ 
  & (0.034) & (0.030) & (0.027) & (0.035) \\ 
  WTP & 0.004$^{**}$ &  &  & 0.002 \\ 
  & (0.002) &  &  & (0.002) \\ 
  Belief &  & 0.003 &  & 0.001 \\ 
  &  & (0.002) &  & (0.002) \\ 
  Gambling Score &  &  & 0.630$^{***}$ & 0.581$^{***}$ \\ 
  &  &  & (0.145) & (0.149) \\ 
 \hline \\[-1.8ex] 
Observations & 285 & 285 & 285 & 285 \\ 
R$^{2}$ & 0.018 & 0.009 & 0.062 & 0.070 \\ 
Adjusted R$^{2}$ & 0.014 & 0.006 & 0.059 & 0.060 \\ 
\hline 
\hline \\[-1.8ex] 
\end{tabular} 
}
\vspace{0.5em}
\begin{minipage}{\textwidth}
\footnotesize
\setlength{\parskip}{0em}
\textit{Notes}: Dependent variable in all columns is a measure for real world overspending on loot boxes, namely a dummy that is equal to one if the subject reports to have spent more than planned on loot boxes. Gambling score is a score from a self reported gambling questionnaire, scaled from 0 to 1. Included are only buyer subjects that know what loot boxes are. \textit{ *~$p<0.1$, **~$p<0.05$, ***~$p<0.01$}
\end{minipage}
\end{table}

\subsection{Discussion: Asymmetric sophistication} One striking aspect of our results is that although buyers and sellers are recruited from the same pool of subjects, sellers appear sophisticated, while buyers seem naive. First of all, we think it adds to the significance of our results that deceptive features are understood and used not only by professional designers of online games, but also by our non-professional seller subjects.   
Second, phenomena related to the apparent ``sellers sophisticated, buyers naive'' dichotomy have been observed in ``information revelation'' studies. For goods with uncertain product quality, for example, the theory suggests that ``no news is bad news'' \citep{milgrom1981good}, meaning that sellers who do not reveal their product quality will be identified by buyers simply because they are not disclosing information. Thus, information should be fully revealed in theory. Empirically, it has been found that information unraveling is incomplete. Buyers often seem not sophisticated enough to understand that withheld information indicates poor quality, and sophisticated sellers exploit this by revealing only positive information. However, the same results can be observed in markets where both buyers and sellers are non-professionals, see \citet{frondel2020power} for an analysis of the property market. In the information unraveling lab experiment of \citet{jin2021nonews}, both parties are represented by human participants, but incomplete unraveling results even when buyers and sellers switch roles. Our results resonate with this literature, confirming the view that asymmetric sophistication drives persistent market distortions.

%% file: Conclusion-Marketing.tex
\section{Conclusion}

\label{sec:Conclusion}

Our paper provides comprehensive evidence on how deceptive seller-buyer interactions lead to overspending on probabilistic product offerings and a reduction in consumer welfare. Deceptive lottery features induce this buyer overspending, so that firms can strategically use deceptive features for revenue maximization. This effect is driven by buyers' overoptimistic beliefs about their chances of winning. Suppliers use such deceptive features strategically to exploit these beliefs. The real-world relevance of our findings is emphasized by their correlation with actual purchasing behavior in the wild. We focus on loot boxes, a common form of deceptively framed lottery that video game developers use to monetize their games, and find that more than 22\% of our buyer subjects reported having spent money on loot boxes within the past 12 months. Moreover, a higher willingness to pay in our experimental setting significantly predicts a greater likelihood of overspending on loot boxes in the real world. These results suggest that, despite the stylized nature of our setup, it captures meaningful aspects of real-world purchasing behavior related to these probabilistic products.

Our findings have direct implications for marketing. The results extend existing work on consumer misperceptions and behavioral pricing by showing how firms can utilize these misperceptions through selective information design. From a managerial perspective, deceptive framing offers a low-cost, scalable means of extracting additional surplus without altering the underlying product. This highlights the economic potential of presentation formats as a marketing instrument, especially in digital markets.

The implications of deceptive framing reach beyond digital gaming. In physical markets, similar mechanisms can be observed, for instance, in the sale of collectible ``blind boxes.'' Pop Mart, a leading producer of such toys with a market capitalization exceeding \$40 billion in 2025, displays equal odds across all items in a series, yet the probability of obtaining the rare ``hidden'' figure is in fact much lower (1 in 144 rather than 1 in 13).\footnote{The true odds are only available in the fine print on the website \url{popmart.com}.} Our findings suggest that such designs operate similarly to those used in digital loot boxes and are, in general, applicable to a wide range of marketing scenarios.

%% file: Appendix_A.tex
\subsection*{Appendix A: Additional Figures and Tables}
\label{sec:Appendix A}

\begin{figure}[!htbp]
    \centering
    \caption{Median offer prices by treatment} 
    \vspace{0.5em}
    \includegraphics[width=0.6\textwidth]{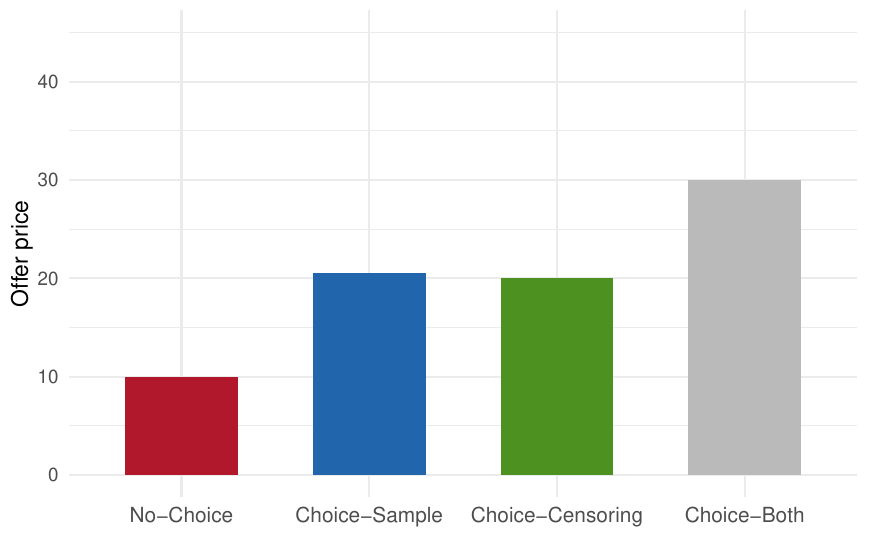}
    \label{fig:median offer}
    \begin{flushleft} 
        \footnotesize
        \setlength{\parskip}{0em} 
        \textit{Note:} Median offer prices in the different treatments. The medians are higher in treatments with deceptive features
    \end{flushleft}
\end{figure}

\begin{table}[!htbp]
\centering
\caption{Consistency of presentation form choices by treatment}
\label{tab:consistency}
\renewcommand{\arraystretch}{1.2}
\begin{tabular}{lr lr lr}
\toprule
\multicolumn{2}{c}{\textbf{Choice-Sample}} & \multicolumn{2}{c}{\textbf{Choice-Censoring}} & \multicolumn{2}{c}{\textbf{Choice-Both}} \\
\textbf{Choice} & \textbf{\%} & \textbf{Choice} & \textbf{\%} & \textbf{Choice} & \textbf{\%} \\
\midrule
Always Sample & 47.9 & Always Censoring & 61.6 & Always Censoring & 3.0 \\
Always Transparent & 4.2 & Always Transparent & 9.1 & Always Sample & 5.0 \\
Mixed choices & 47.9 & Mixed choices & 29.3 & Always Sample\_Censoring & 10.0 \\
 &  &  &  & Always Transparent & 3.0 \\
 &  &  &  & Mixed choices & 79.0 \\
\bottomrule
\end{tabular}
\begin{flushleft}
\footnotesize
\setlength{\parskip}{0em}
\textit{Notes}: Consistency of choices in the treatments with deceptive features. In the treatment \textit{Choice-Both} 79\% of subjects used multiple presentation forms in the five rounds.
\end{flushleft}
\end{table}

\begin{table}[!htbp] \centering 
  \caption{Frequency of categories in which sellers justified their choices} 
  \label{tab:justification_categories} 
  \resizebox{\textwidth}{!}{
\begin{tabular}{@{\extracolsep{5pt}} lcccc} 
\\[-1.8ex]\hline 
\hline \\[-1.8ex] 
& \multicolumn{4}{c}{Category} \\
\cline{2-5} 
\\[-1.8ex]
 & Perceived Value & Simplicity and Clarity & Emotional Appeal & Other \\
 & (1) & (2) & (3) & (4) \\
\hline \\[-1.8ex] 
All & $163\,(55.3\%)$ & $71\,(24.1\%)$ & $15\,(5.1\%)$ & $46\,(15.6\%)$ \\
Censoring & $56\,(65.1\%)$ & $12\,(14.0\%)$ & $5\,(5.8\%)$ & $13\,(15.1\%)$ \\
Sample & $60\,(59.4\%)$ & $18\,(17.8\%)$ & $8\,(7.9\%)$ & $15\,(14.9\%)$ \\
Sample\_Censoring & $34\,(65.4\%)$ & $8\,(15.4\%)$ & $1\,(1.9\%)$ & $9\,(17.3\%)$ \\
Transparent & $13\,(23.2\%)$ & $33\,(58.9\%)$ & $1\,(1.8\%)$ & $9\,(16.1\%)$ \\
\hline
\hline \\[-1.8ex] 
\end{tabular} 
}
\vspace{0.5em}
\begin{minipage}{\textwidth}
\footnotesize
\setlength{\parskip}{0em}
\textit{Notes}: We focus on the free-text justifications of the seller's choices in round 1. We use GPT-4o-mini with temperature $0$ to elicit four categories: \textit{Perceived Value}, \textit{Simplicity \& Clarity}, \textit{Emotional Appeal}, and \textit{Other}. Each justification was then classified ten times with temperature $0.7$. The temperature parameter controls randomness in the model’s output, with $0$ producing answers closer to being deterministic. Ambiguous cases were coded as \textit{Other}. Sellers that pick deceptive features cite \textit{Perceived Value} more often than Transparent, while Transparent choices emphasize \textit{Simplicity \& Clarity}.
\end{minipage}
\end{table}

\begin{figure}[!htbp]
    \centering
    \caption{Mean second-order belief by presentation form with 95\% CI} 
    \vspace{0.5em}
    \includegraphics[width=0.6\textwidth]{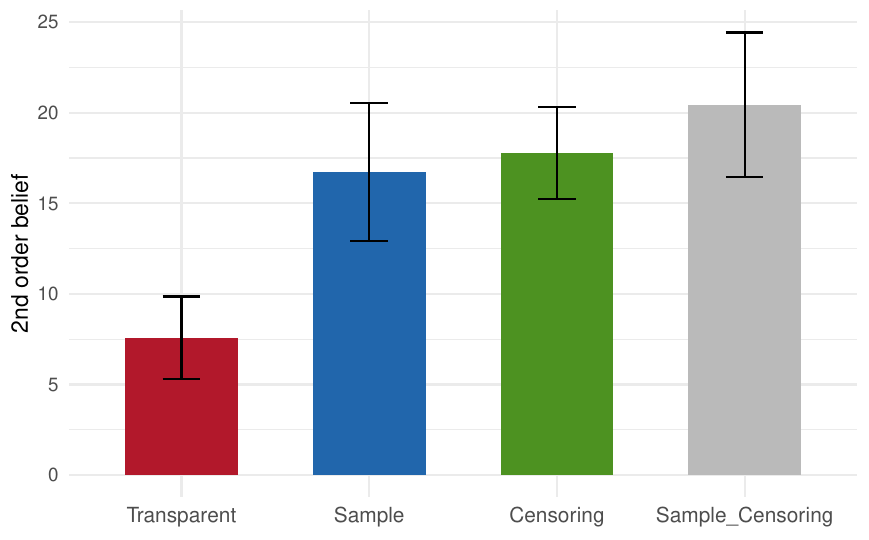}
    \label{fig:belief seller presentationform}
    \begin{flushleft} 
        \footnotesize
        \setlength{\parskip}{0em} 
        Mean second-order beliefs by the presentation form that the sellers have chosen. Whiskers are the 95\% CI.
    \end{flushleft}
\end{figure}

\begin{table}[!htbp] \centering 
  \caption{Predictors for real world loot box spending - regression results} 
  \label{tab:loot box spending} 
 { \small
\begin{tabular}{@{\extracolsep{5pt}}lcccc} 
\\[-1.8ex]\hline 
\hline \\[-1.8ex] 
 & \multicolumn{4}{c}{Loot box spending} \\ 
\cline{2-5} 
\\[-1.8ex] & (1) & (2) & (3) & (4)\\ 
\hline \\[-1.8ex] 
 Constant & $-$0.506 & 11.767$^{**}$ & 8.547$^{*}$ & $-$2.141 \\ 
  & (5.977) & (5.375) & (4.819) & (6.261) \\ 
  WTP & 0.999$^{***}$ &  &  & 1.012$^{***}$ \\ 
  & (0.281) &  &  & (0.315) \\ 
  Belief &  & 0.257 &  & $-$0.307 \\ 
  &  & (0.292) &  & (0.319) \\ 
  Gambling Score &  &  & 69.243$^{***}$ & 55.308$^{**}$ \\ 
  &  &  & (26.337) & (26.632) \\ 
 \hline \\[-1.8ex] 
Observations & 285 & 285 & 285 & 285 \\ 
R$^{2}$ & 0.043 & 0.003 & 0.024 & 0.059 \\ 
Adjusted R$^{2}$ & 0.039 & $-$0.001 & 0.020 & 0.049 \\ 
\hline 
\hline \\[-1.8ex] 
\end{tabular} 
}
\vspace{0.5em}
\begin{minipage}{\textwidth}
\footnotesize
\setlength{\parskip}{0em}
\textit{Notes}: Dependent variable is the real world loot box spending.  \textit{ *~$p<0.1$, **~$p<0.05$, ***~$p<0.01$}
\end{minipage}
\end{table}

%% file: Appendix_B.tex
\subsection*{Appendix B: Instructions}
\label{sec:Appendix B}

Below we provide screenshots of all pages of the experiment. Included are attention and comprehension checks, instructions, and all survey questions. The screenshots are presented in the order in which participants progress through the experiment. We first start with the seller stage.

\begin{figure}[!htbp]
    \centering
    \caption{Welcome page - Seller} 
    \vspace{0.5em}
    \includegraphics[width=0.7\textwidth]{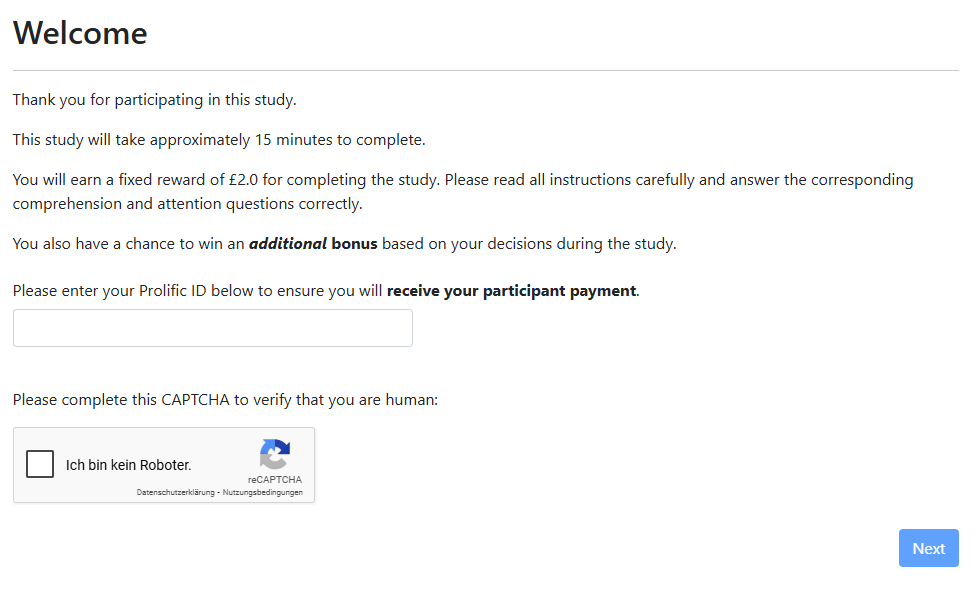}
    \label{fig:welcome seller}
    \begin{flushleft} 
        \footnotesize
        \setlength{\parskip}{0em} 
       Welcome page of the experiment. A CAPTCHA ensures that only humans can progress in the experiment. 
    \end{flushleft}
\end{figure}

\begin{figure}[!htbp]
    \centering
    \caption{Attention check page - Seller} 
    \vspace{0.5em}
    \includegraphics[width=0.7\textwidth]{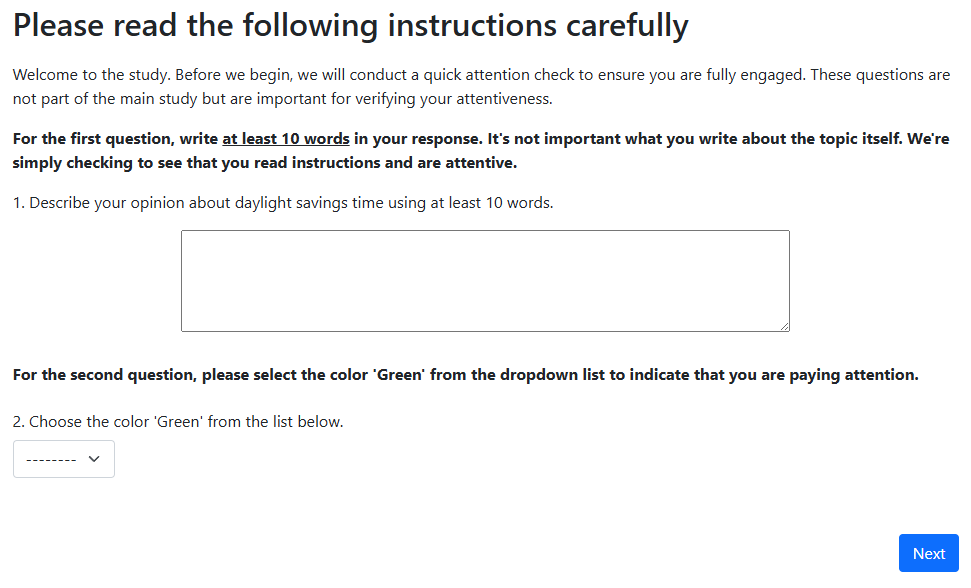}
    \label{fig:attention seller}
    \begin{flushleft} 
        \footnotesize
        \setlength{\parskip}{0em} 
       Attention check page of the experiment. Subjects that failed the attention check were redirected to Prolific. 
    \end{flushleft}
\end{figure}

\begin{figure}[!htbp]
    \centering
    \caption{Instructions treatment \textit{Transparent} - Seller} 
    \vspace{0.5em}
    \includegraphics[width=1\textwidth]{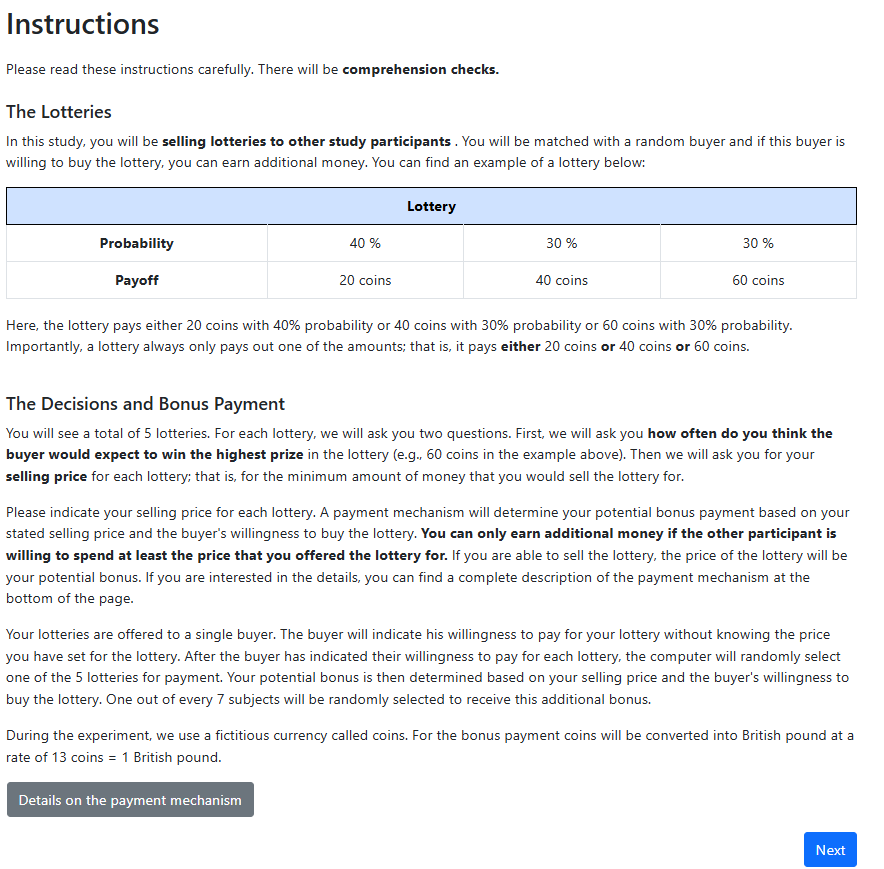}
    \label{fig:instructions transparent seller}
\end{figure}

\begin{figure}[!htbp]
    \centering
    \caption{Instructions treatments with deceptive features - Seller} 
    \vspace{0.5em}
    \includegraphics[width=1\textwidth]{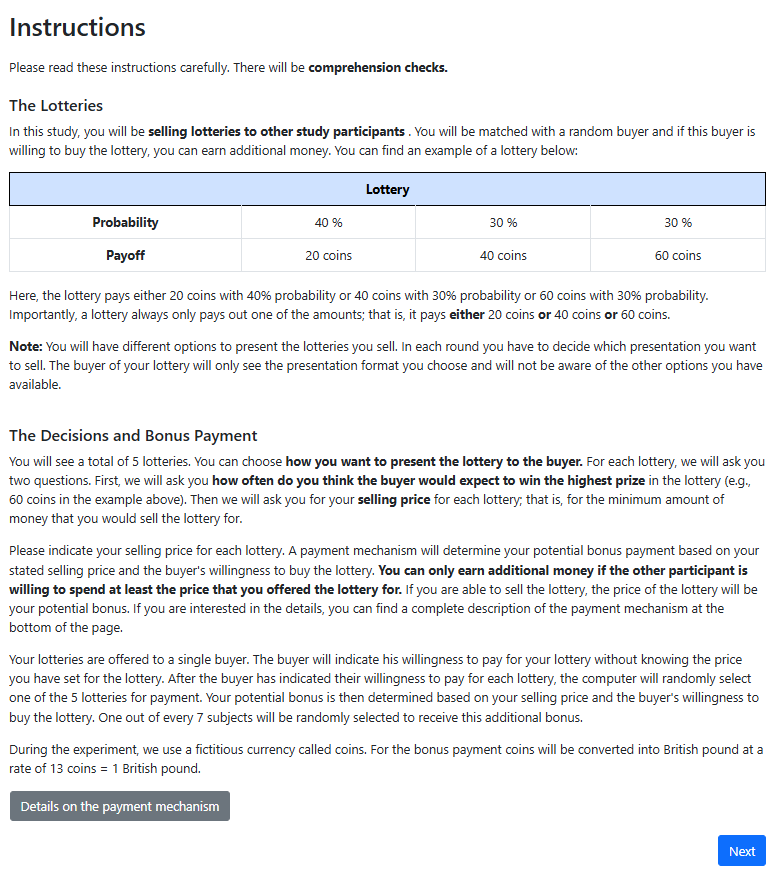}
    \label{fig:instructions deceptive feature seller}
    \begin{flushleft} 
        \footnotesize
        \setlength{\parskip}{0em} 
       In contrast to the instructions for the treatment \textit{Transparent}, seller subjects are informed that they can choose the presentation of the lottery in the treatments with deceptive features. 
    \end{flushleft}
\end{figure}

\begin{figure}[!htbp]
    \centering
    \caption{Payment details box - Seller} 
    \vspace{0.5em}
    \includegraphics[width=1\textwidth]{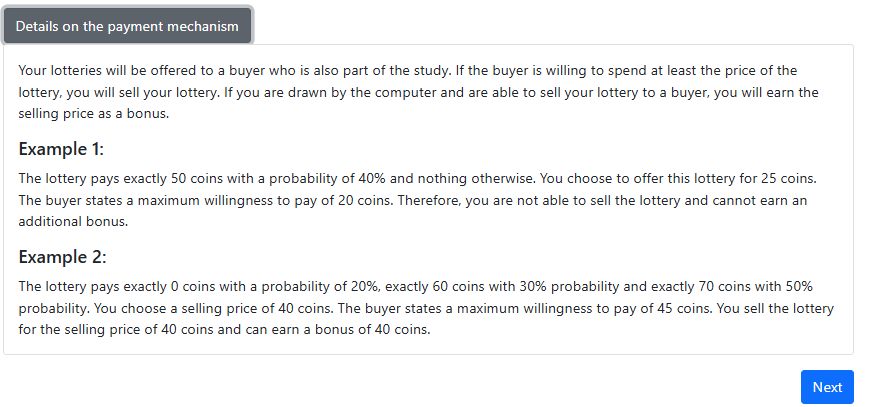}
    \label{fig:payment details seller}
    \begin{flushleft} 
        \footnotesize
        \setlength{\parskip}{0em} 
    \end{flushleft}
\end{figure}

\begin{figure}[!htbp]
    \centering
    \caption{Comprehension checks - Seller} 
    \vspace{0.5em}
    \includegraphics[width=0.7\textwidth]{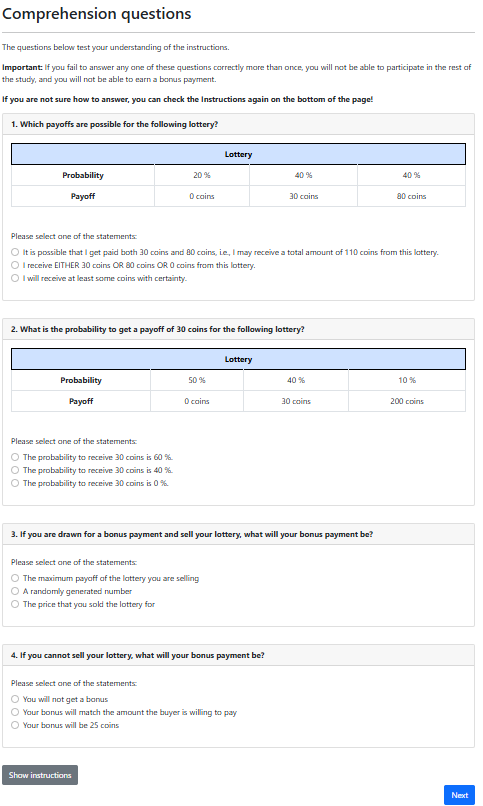}
    \label{fig:comprehension seller}
    \begin{flushleft} 
        \footnotesize
        \setlength{\parskip}{0em} 
       Subjects had to answer four comprehension questions. They were given to tries to correctly answer all comprehension questions. Instructions could be opened in an additional box. Subjects that failed the comprehension questions more than once were asked to return the study on Prolific.
    \end{flushleft}
\end{figure}

\begin{figure}[!htbp]
    \centering
    \caption{Start experiment - Seller} 
    \vspace{0.5em}
    \includegraphics[width=0.9\textwidth]{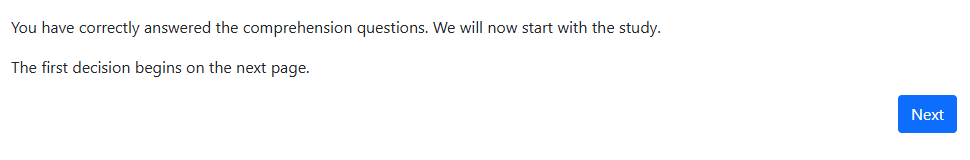}
    \label{fig:start seller}
\end{figure}

\begin{figure}[!htbp]
    \centering
    \caption{Presentation decision \textit{Censoring} - Seller} 
    \vspace{0.5em}
    \includegraphics[width=0.9\textwidth]{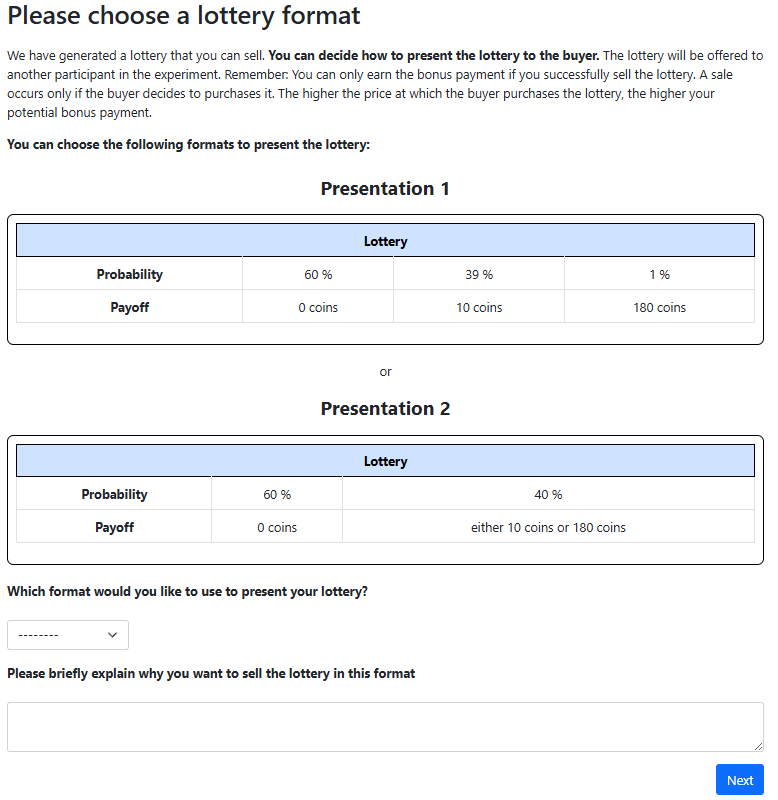}
    \label{fig:decision censoring}
\end{figure}

\begin{figure}[!htbp]
    \centering
    \caption{Presentation decision \textit{Sample} - Seller} 
    \vspace{0.5em}
    \includegraphics[width=1\textwidth]{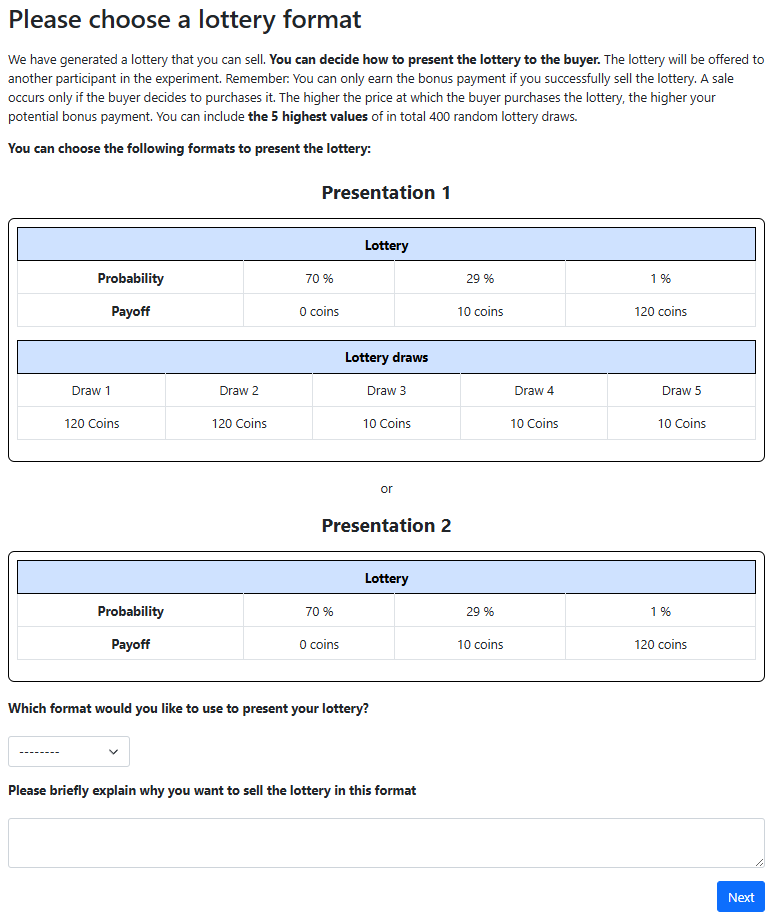}
    \label{fig:decision sample}
\end{figure}

\begin{figure}[!htbp]
    \centering
    \caption{Presentation decision \textit{Sample\_Censoring} - Seller} 
    \vspace{0.5em}
    \includegraphics[width=0.69\textwidth]{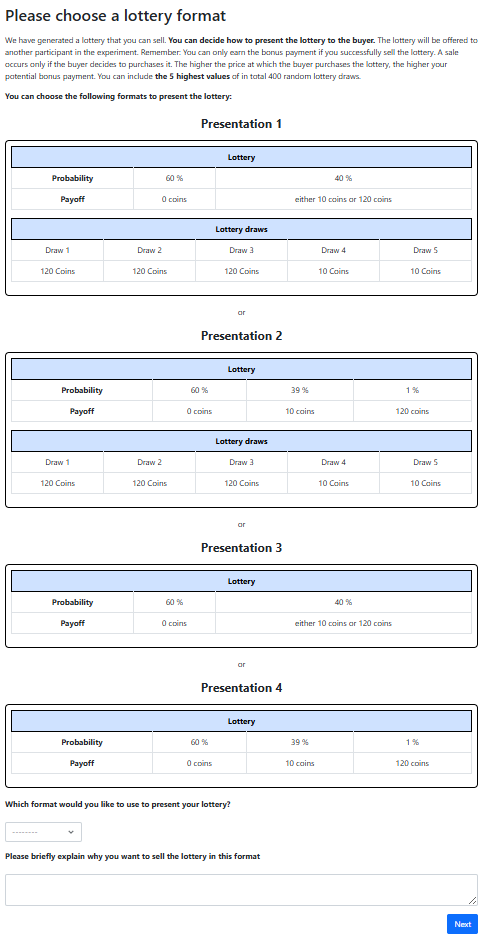}
    \label{fig:decision samplecensoring}
\end{figure}

\begin{figure}[!htbp]
    \centering
    \caption{Price decision - Seller} 
    \vspace{0.5em}
    \includegraphics[width=1\textwidth]{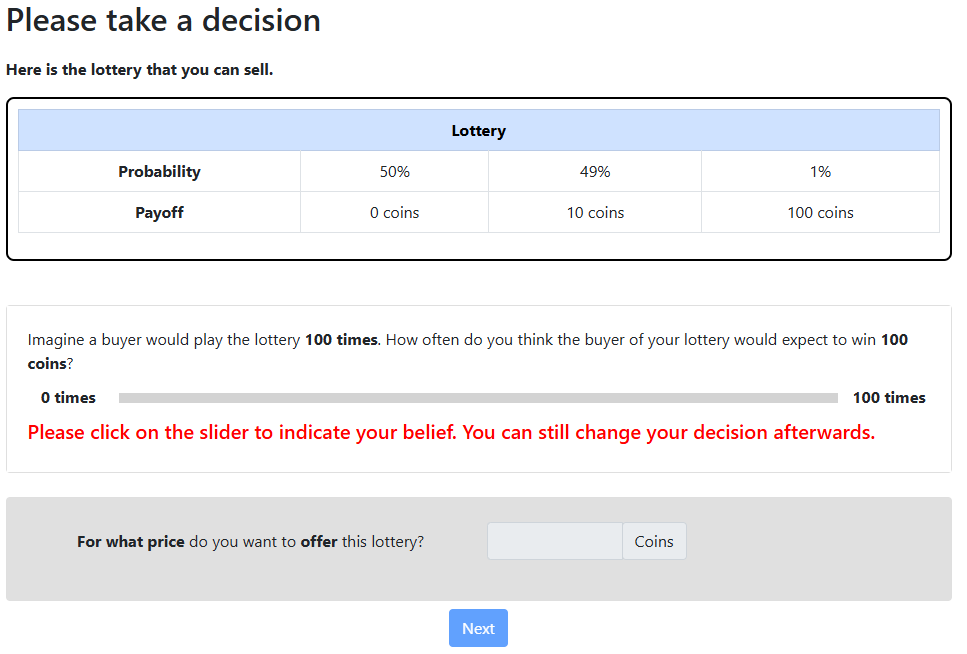}
    \label{fig:price decision seller}
    \begin{flushleft} 
        \footnotesize
        \setlength{\parskip}{0em} 
       The lottery is the presentation form chosen on the previous page (or in \textit{Transparent} the lottery provided by us). The slider for beliefs has no default to avoid anchoring effects. A price can be chosen after the belief is indicated. There are in total five rounds of presentation and price decisions.
    \end{flushleft}
\end{figure}

\begin{figure}[!htbp]
    \centering
    \caption{Demographics - Seller} 
    \vspace{0.5em}
    \includegraphics[width=1\textwidth]{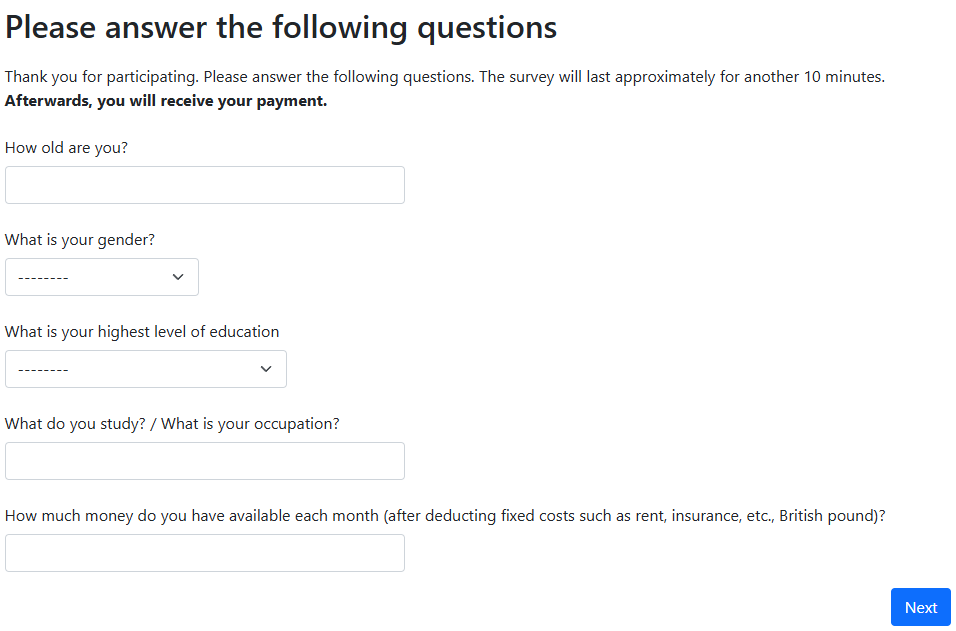}
    \label{fig:demographics seller}
\end{figure}

\begin{figure}[!htbp]
    \centering
    \caption{Loot box questions - Seller} 
    \vspace{0.5em}
    \includegraphics[width=0.8\textwidth]{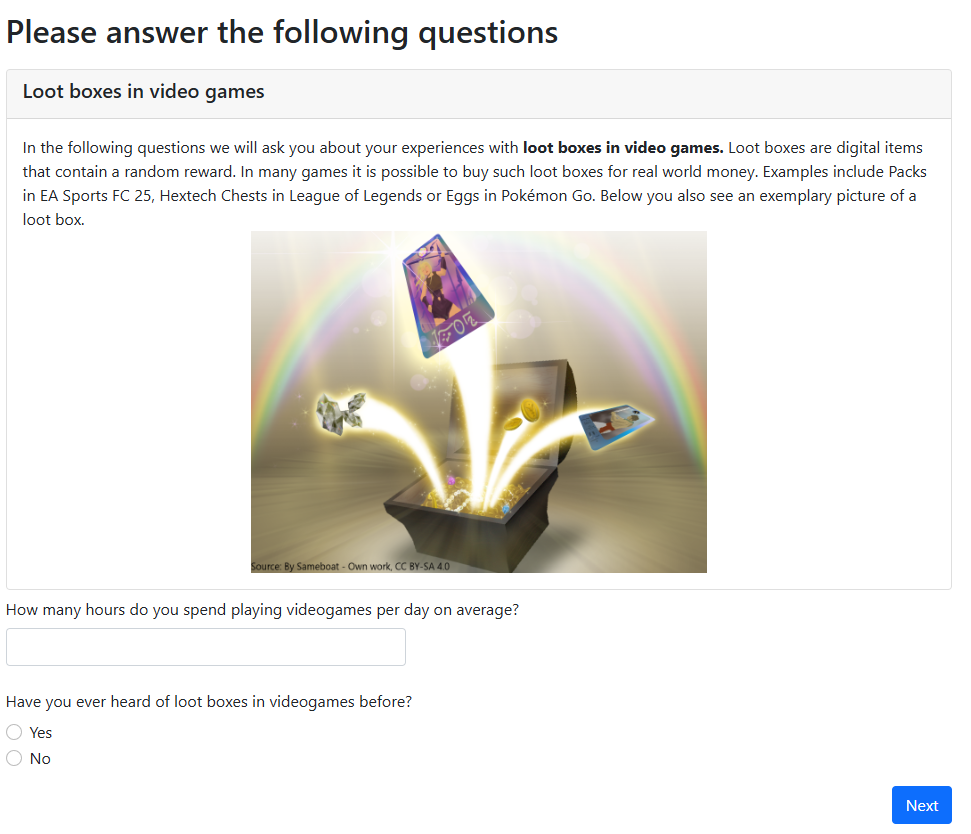}
    \label{fig:loot box seller}
\end{figure}

\begin{figure}[!htbp]
    \centering
    \caption{Loot box spending - Seller} 
    \vspace{0.5em}
    \includegraphics[width=0.8\textwidth]{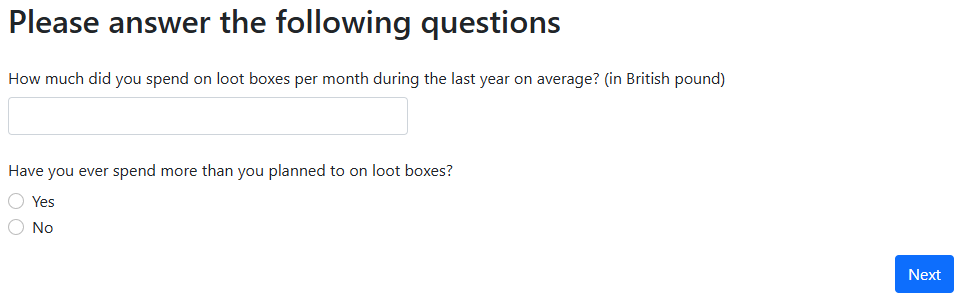}
    \label{fig:loot box yes seller}
    \begin{flushleft} 
        \footnotesize
        \setlength{\parskip}{0em} 
       If subjects had heard of loot boxes before, we ask them about their loot box spending.
    \end{flushleft}
\end{figure}

\begin{figure}[!htbp]
    \centering
    \caption{Self control questions - Seller} 
    \vspace{0.5em}
    \includegraphics[width=1\textwidth]{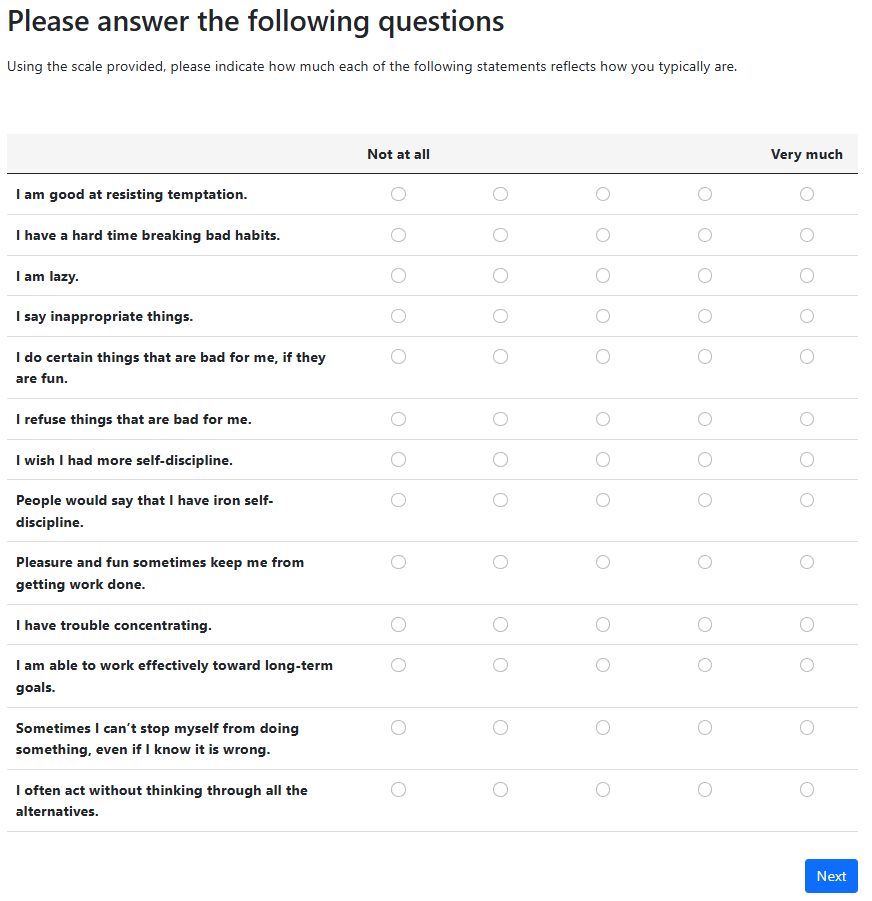}
    \label{fig:self control seller}
\end{figure}

\begin{figure}[!htbp]
    \centering
    \caption{Gambling behavior - Seller} 
    \vspace{0.5em}
    \includegraphics[width=1\textwidth]{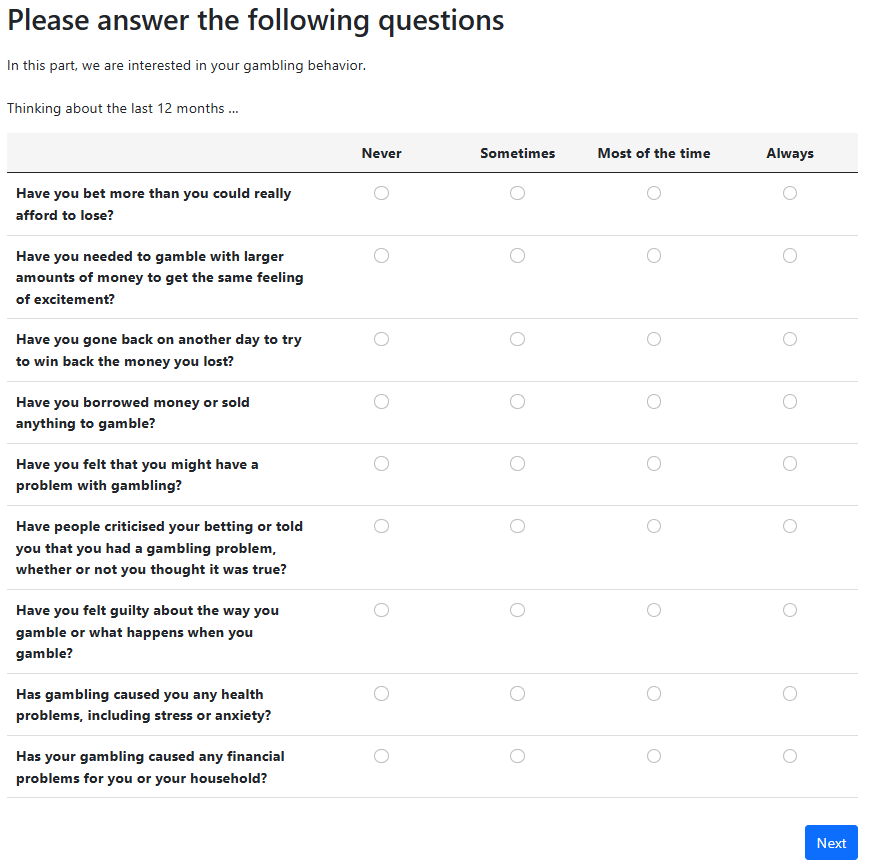}
    \label{fig:gambling seller}
\end{figure}

\begin{figure}[!htbp]
    \centering
    \caption{Debriefing - Seller} 
    \vspace{0.5em}
    \includegraphics[width=1\textwidth]{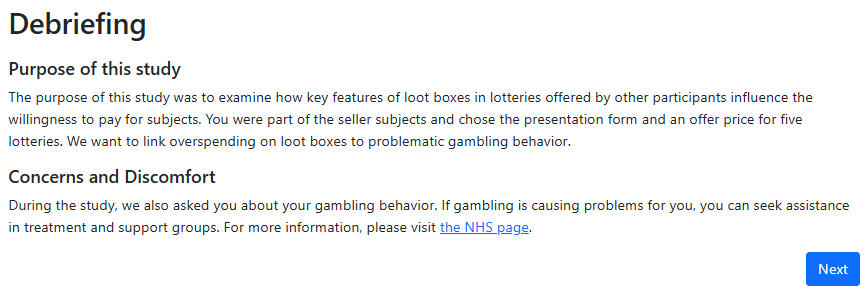}
    \label{fig:debriefing seller}
\end{figure}

\begin{figure}[!htbp]
    \centering
    \caption{Final page - Seller} 
    \vspace{0.5em}
    \includegraphics[width=1\textwidth]{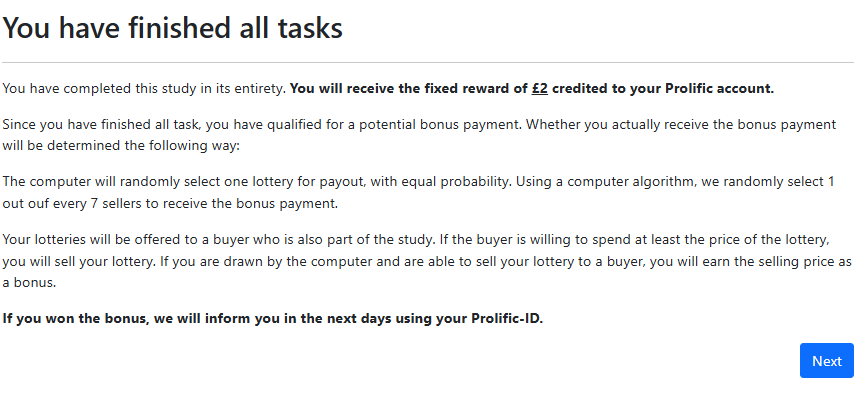}
    \label{fig:final seller}
\end{figure}

\begin{figure}[!htbp]
    \centering
    \caption{Welcome page - Buyer} 
    \vspace{0.5em}
    \includegraphics[width=0.9\textwidth]{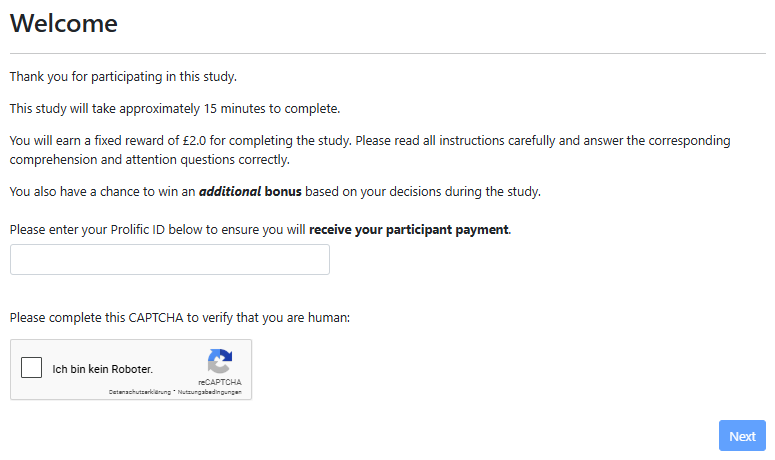}
    \label{fig:welcome buyer}
    \begin{flushleft} 
        \footnotesize
        \setlength{\parskip}{0em} 
       Welcome page of the experiment. A CAPTCHA ensures that only humans can progress in the experiment. 
    \end{flushleft}
\end{figure}

\begin{figure}[!htbp]
    \centering
    \caption{Attention check page - Buyer} 
    \vspace{0.5em}
    \includegraphics[width=0.9\textwidth]{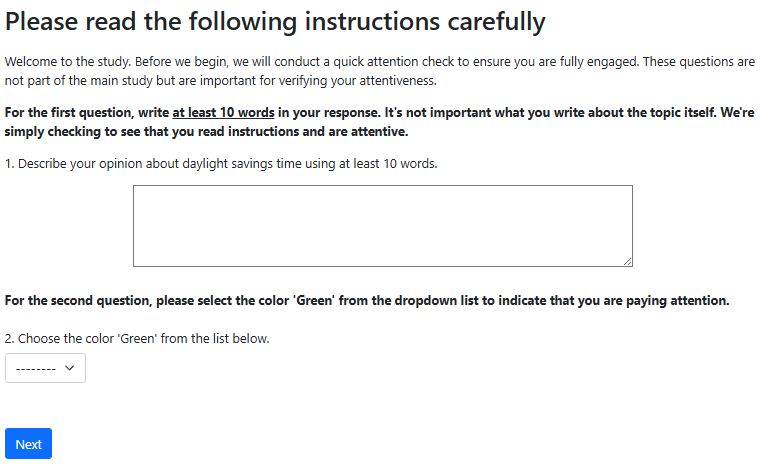}
    \label{fig:attention buyer}
    \begin{flushleft} 
        \footnotesize
        \setlength{\parskip}{0em} 
       Attention check page of the experiment. Subjects that failed the attention check were redirected to Prolific. 
    \end{flushleft}
\end{figure}

\begin{figure}[!htbp]
    \centering
    \caption{Instructions - Buyer} 
    \vspace{0.5em}
    \includegraphics[width=0.8\textwidth]{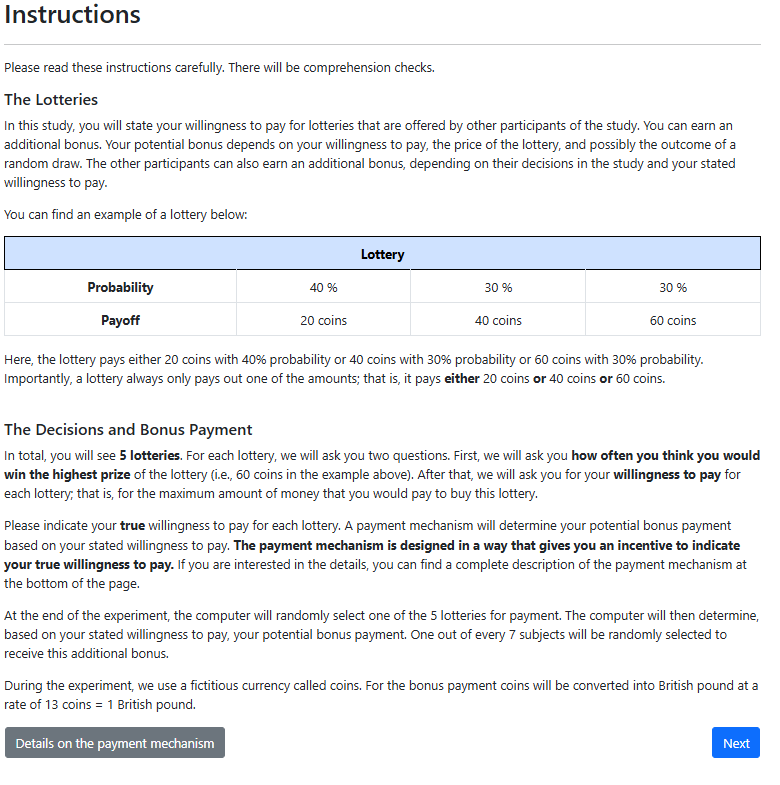}
    \label{fig:instructions buyer}
\end{figure}

\begin{figure}[!htbp]
    \centering
    \caption{Payment details box - Buyer} 
    \vspace{0.5em}
    \includegraphics[width=0.8\textwidth]{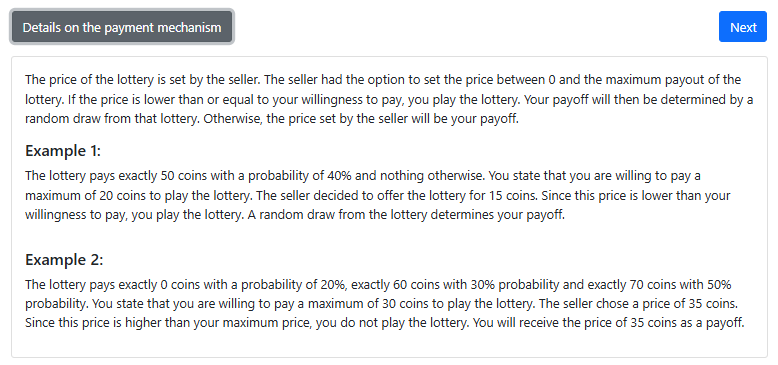}
    \label{fig:payment details buyer}
\end{figure}

\begin{figure}[!htbp]
    \centering
    \caption{Comprehension checks - Buyer} 
    \vspace{0.5em}
    \includegraphics[width=0.85\textwidth]{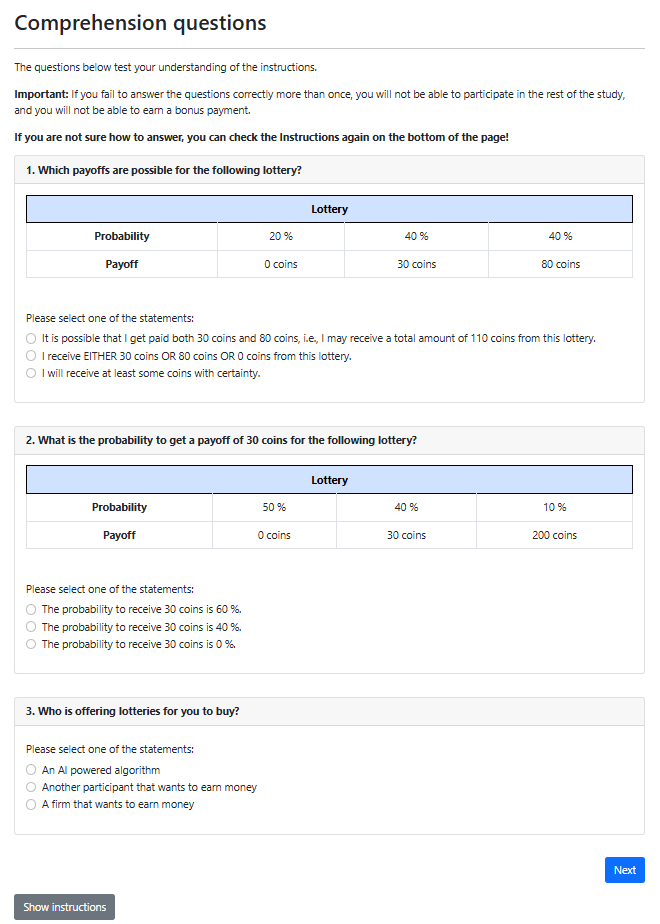}
    \label{fig:comprehension buyer}
    \begin{flushleft} 
        \footnotesize
        \setlength{\parskip}{0em} 
       Subjects had to answer three comprehension questions. They were given to tries to correctly answer all comprehension questions. Instructions could be opened in an additional box. Subjects that failed the comprehension questions more than once were asked to return the study on Prolific.
    \end{flushleft}
\end{figure}

\begin{figure}[!htbp]
    \centering
    \caption{Start experiment - Buyer} 
    \vspace{0.5em}
    \includegraphics[width=1\textwidth]{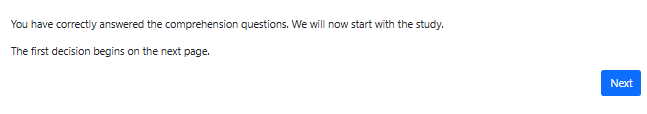}
    \label{fig:start buyer}
\end{figure}

\begin{figure}[!htbp]
    \centering
    \caption{WTP decision - Buyer} 
    \vspace{0.5em}
    \includegraphics[width=1\textwidth]{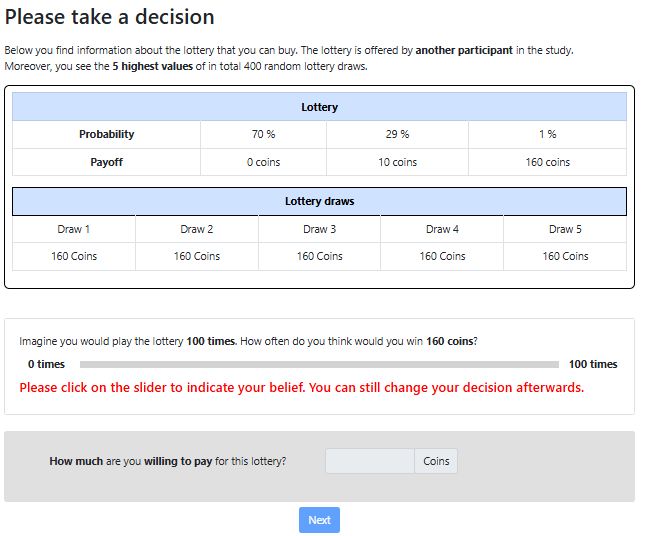}
    \label{fig:decision wtp}
    \begin{flushleft} 
        \footnotesize
        \setlength{\parskip}{0em} 
       The slider for beliefs has no default to avoid anchoring effects. A price can be chosen after the belief is indicated.
    \end{flushleft}
\end{figure}

\begin{figure}[!htbp]
    \centering
    \caption{Feedback lottery bought - Buyer} 
    \vspace{0.5em}
    \includegraphics[width=0.9\textwidth]{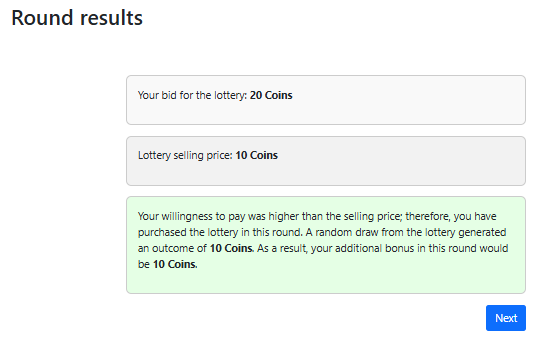}
    \label{fig:feedback bought}
\end{figure}

\begin{figure}[!htbp]
    \centering
    \caption{Feedback lottery not bought - Buyer} 
    \vspace{0.5em}
    \includegraphics[width=0.9\textwidth]{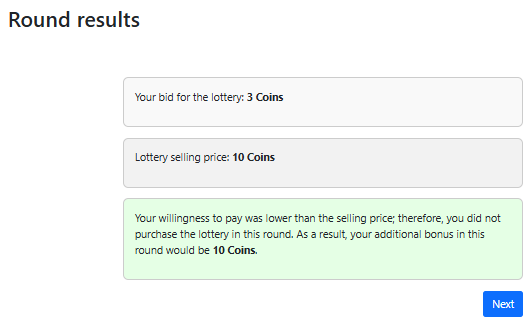}
    \label{fig:feedback not bought}
\end{figure}

\begin{figure}[!htbp]
    \centering
    \caption{Demographics - Buyer} 
    \vspace{0.5em}
    \includegraphics[width=1\textwidth]{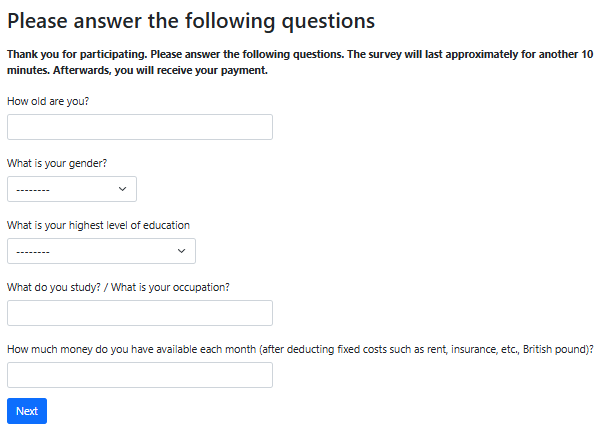}
    \label{fig:demographics buyer}
\end{figure}

\begin{figure}[!htbp]
    \centering
    \caption{Loot box questions - Buyer} 
    \vspace{0.5em}
    \includegraphics[width=0.8\textwidth]{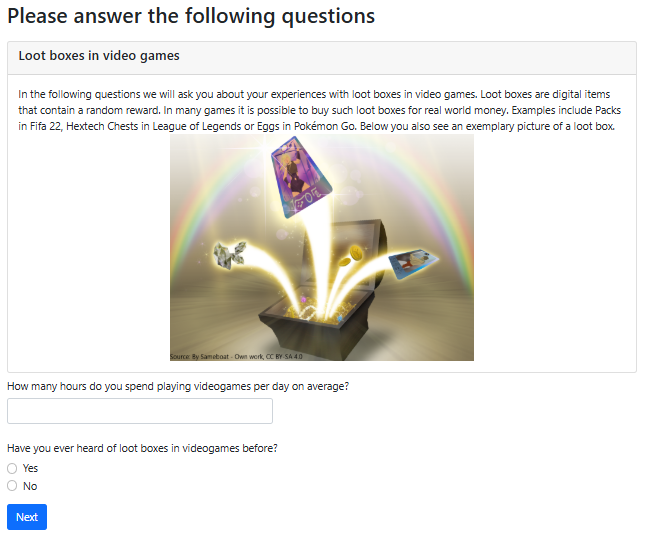}
    \label{fig:loot box buyer}
\end{figure}

\begin{figure}[!htbp]
    \centering
    \caption{Loot box spending - Buyer} 
    \vspace{0.5em}
    \includegraphics[width=0.8\textwidth]{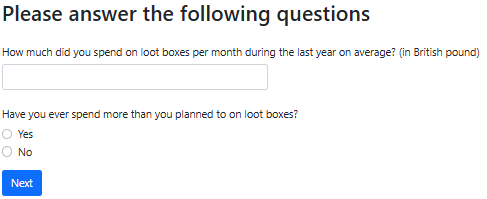}
    \label{fig:loot box yes buyer}
    \begin{flushleft} 
        \footnotesize
        \setlength{\parskip}{0em} 
       If subjects had heard of loot boxes before, we ask them about their loot box spending.
    \end{flushleft}
\end{figure}

\begin{figure}[!htbp]
    \centering
    \caption{Self control questions - Buyer} 
    \vspace{0.5em}
    \includegraphics[width=1\textwidth]{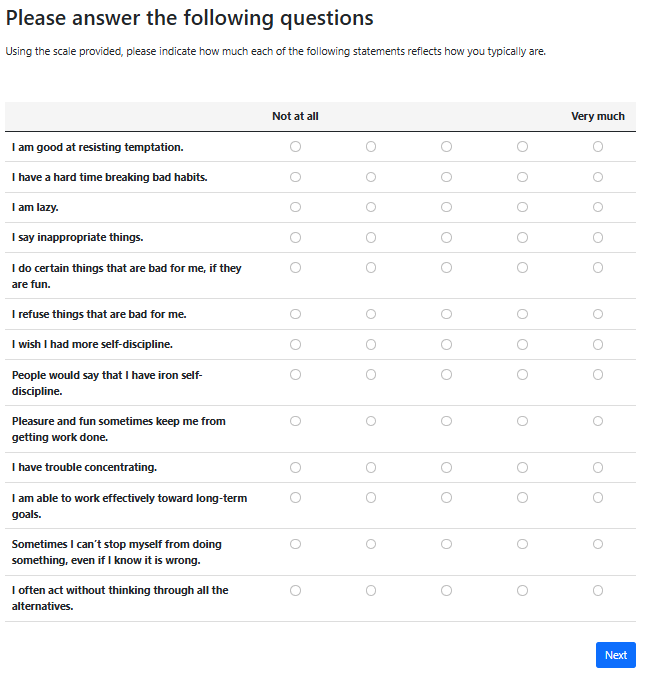}
    \label{fig:self control buyer}
\end{figure}

\begin{figure}[!htbp]
    \centering
    \caption{Gambling behavior - Buyer} 
    \vspace{0.5em}
    \includegraphics[width=1\textwidth]{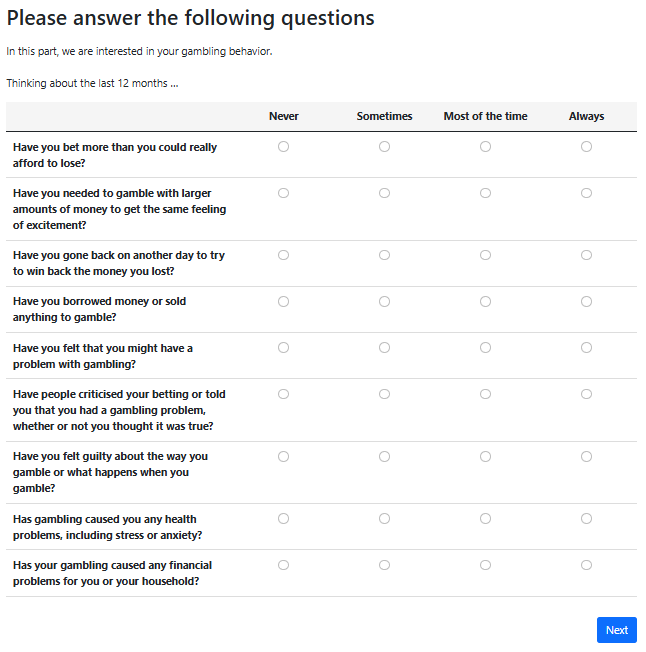}
    \label{fig:gambling buyer}
\end{figure}

\begin{figure}[!htbp]
    \centering
    \caption{Debriefing - Buyer} 
    \vspace{0.5em}
    \includegraphics[width=1\textwidth]{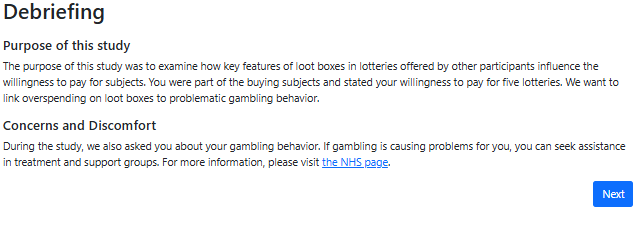}
    \label{fig:debriefing buyer}
\end{figure}

\begin{figure}[!htbp]
    \centering
    \caption{Final page - Buyer} 
    \vspace{0.5em}
    \includegraphics[width=1\textwidth]{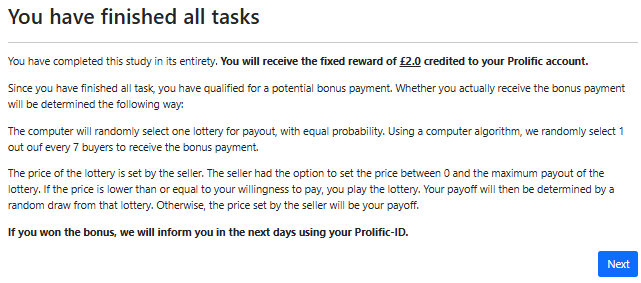}
    \label{fig:final buyer}
\end{figure}

%% file: Appendix_C.tex
\subsection*{Appendix C: Examples for deceptive features in loot boxes}
\label{sec:Appendix C}

Below, we provide examples of the use of deceptive features in video games. 

\begin{figure}[!htbp]
    \centering
    \caption{EA SPORTS FC (formerly FIFA) - Deceptive feature: Censored odds} 
    \vspace{0.5em}
    \includegraphics[width=0.6\textwidth]{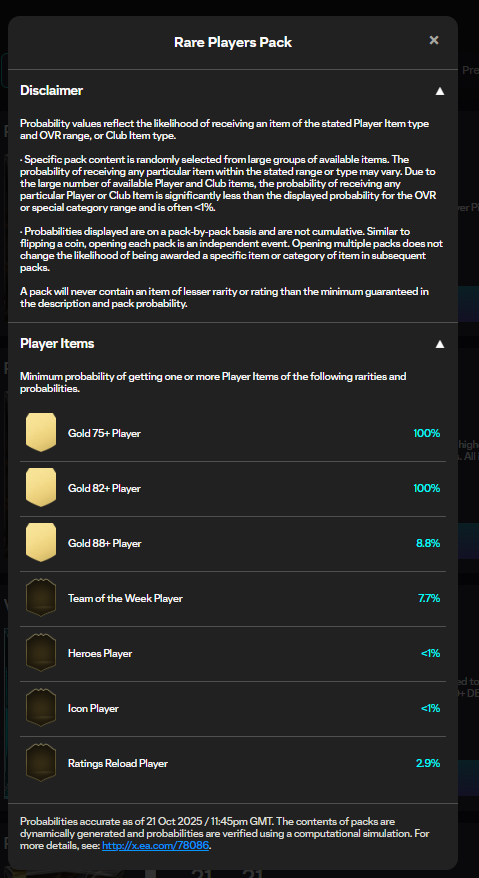}
    \label{fig:Censored EA FC}
    \begin{flushleft} 
        \footnotesize
        \setlength{\parskip}{0em} 
       In \textit{EA Sports FC} odds are communicated for intervals of player strength. 
    \end{flushleft}
\end{figure}

\begin{figure}[!htbp]
    \centering
    \caption{Counter-Strike} 
    \vspace{0.5em}
    \includegraphics[width=0.45\textwidth]{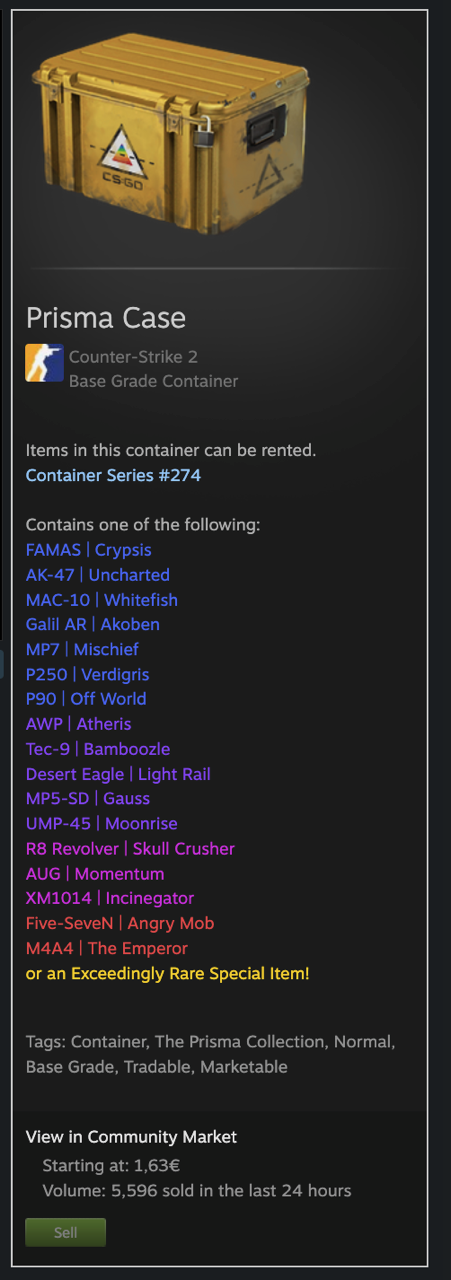}
    \label{fig:Counterstrike}
    \begin{flushleft} 
        \footnotesize
        \setlength{\parskip}{0em} 
       In \textit{Counter-Strike} odds are not directly communicated at the point of sale. Due to regulation, the odds for rarity have been disclosed in the Chinese market.
    \end{flushleft}
\end{figure}

\begin{figure}[!htbp]
\centering
\caption{Pokemon TCGP - Deceptive feature: Selective feedback}
\vspace{0.5em} 
\begin{subfigure}[t]{0.48\textwidth}
    \centering
    \includegraphics[width=\textwidth]{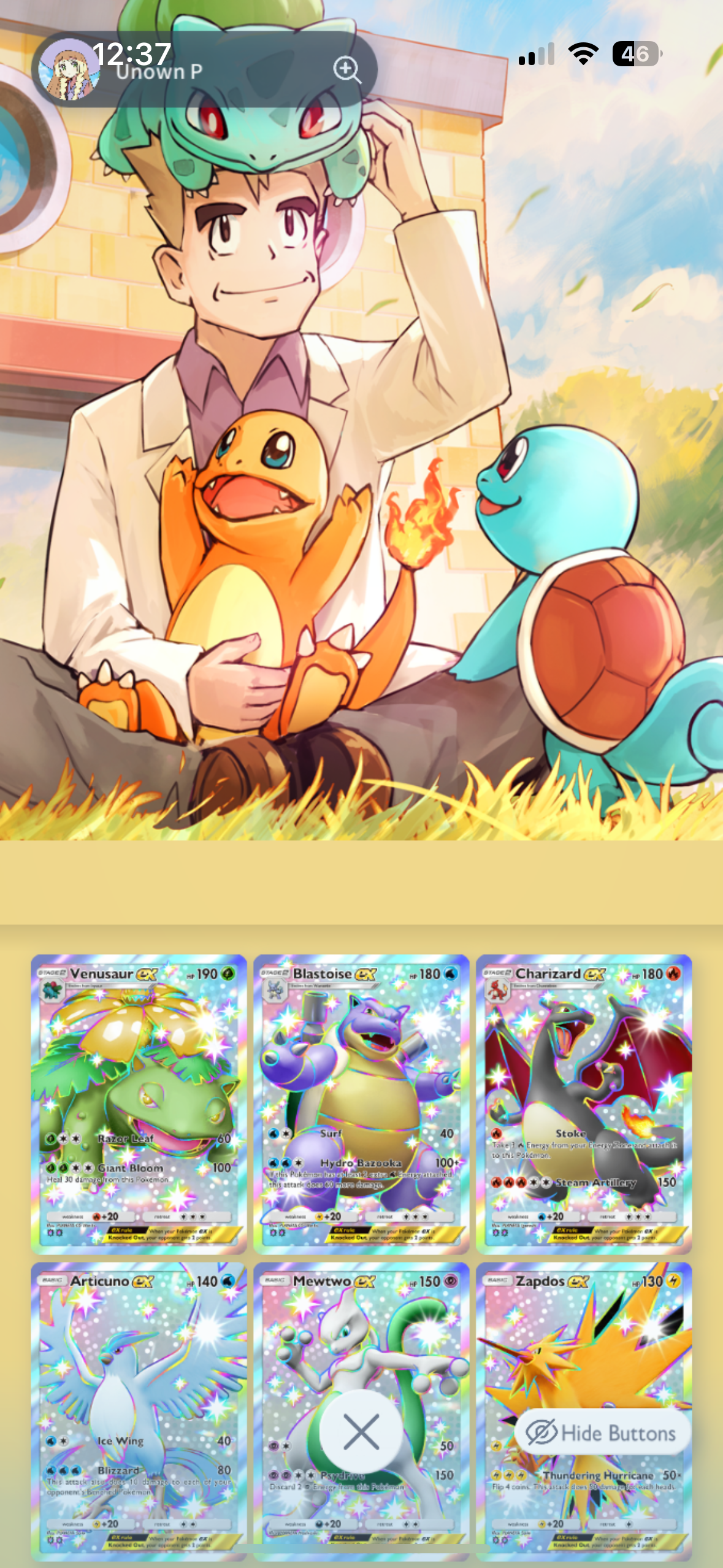}
\end{subfigure}
\hfill
\begin{subfigure}[t]{0.48\textwidth}
    \centering
    \includegraphics[width=\textwidth]{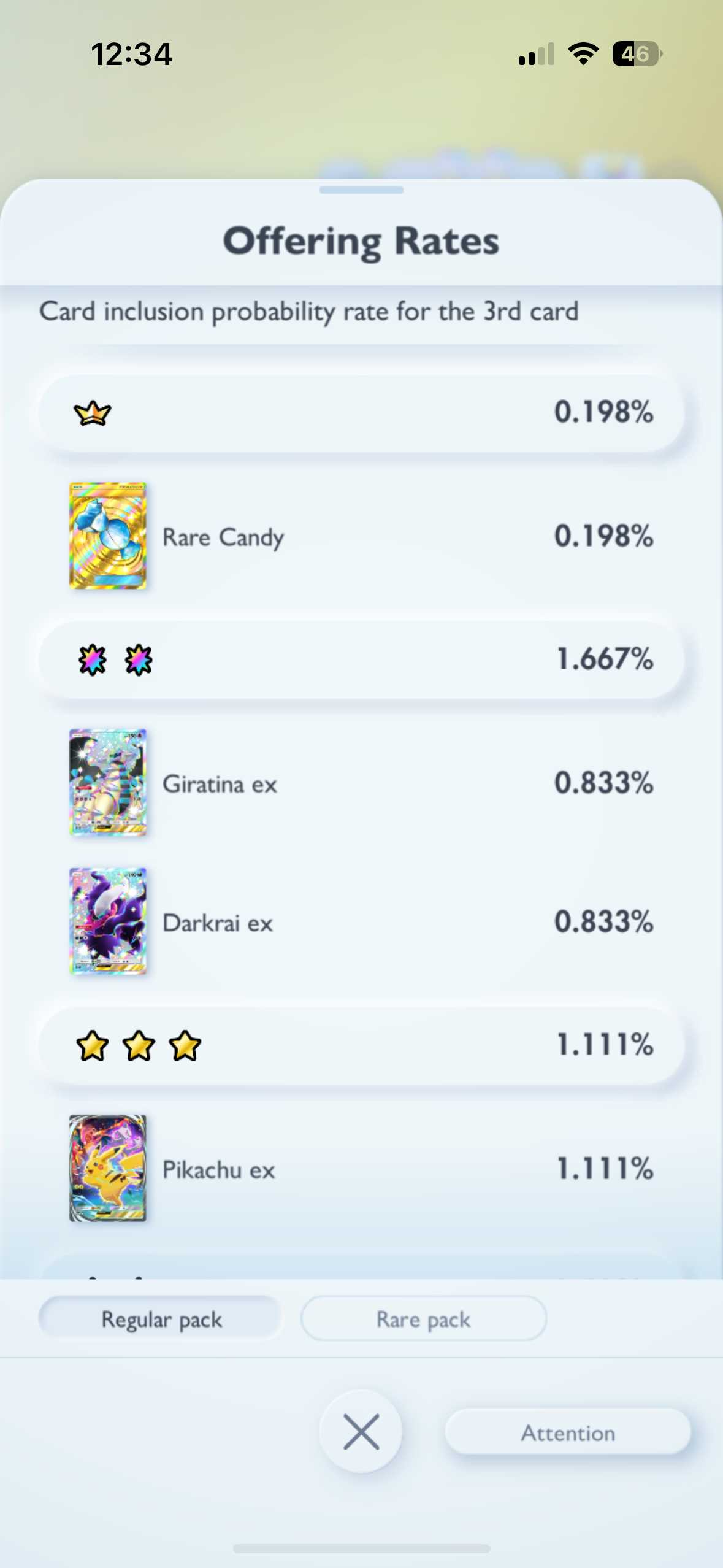}
\end{subfigure}
\label{fig:Pokemon TCGP}
\begin{flushleft} 
        \footnotesize
        \setlength{\parskip}{0em} 
        The left panel shows an example for a community showcase in the mobile game \textit{Pokemon TCGP}. In those, gamers can display cards of their choice. Showcases with rarer cards get more likes and are displayed more prominently. On the right, the odds communication is shown. In \textit{Pokemon TCGP} the odds are communicated transparently.
    \end{flushleft}
\end{figure}

\begin{figure}[!htbp]
    \centering
    \caption{Clash Royale - Deceptive feature: Censored odds} 
    \vspace{0.5em}
    \includegraphics[width=0.45\textwidth]{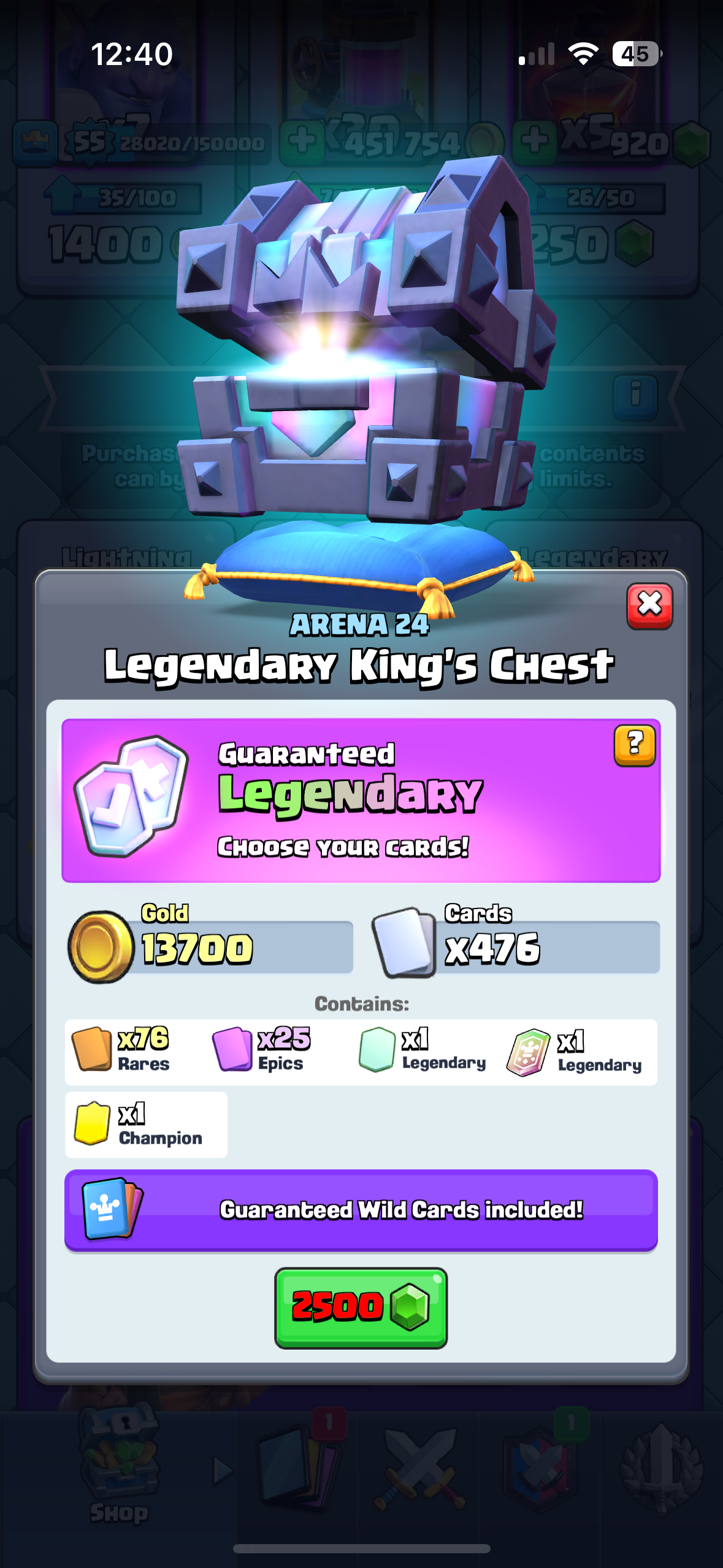}
    \label{fig:Clash Royale}
    \begin{flushleft} 
        \footnotesize
        \setlength{\parskip}{0em} 
       In the mobile game \textit{Clash Royale}, information about how many cards from each rarity tier a loot box contains is shared in the store. It is not communicated what the odds to obtain a specific item from a rarity tier.
    \end{flushleft}
\end{figure}

\begin{figure}[!htbp]
    \centering
    \caption{Raid: Shadow Legends - Deceptive feature: Selective feedback} 
    \vspace{0.5em}
    \includegraphics[width=1\textwidth]{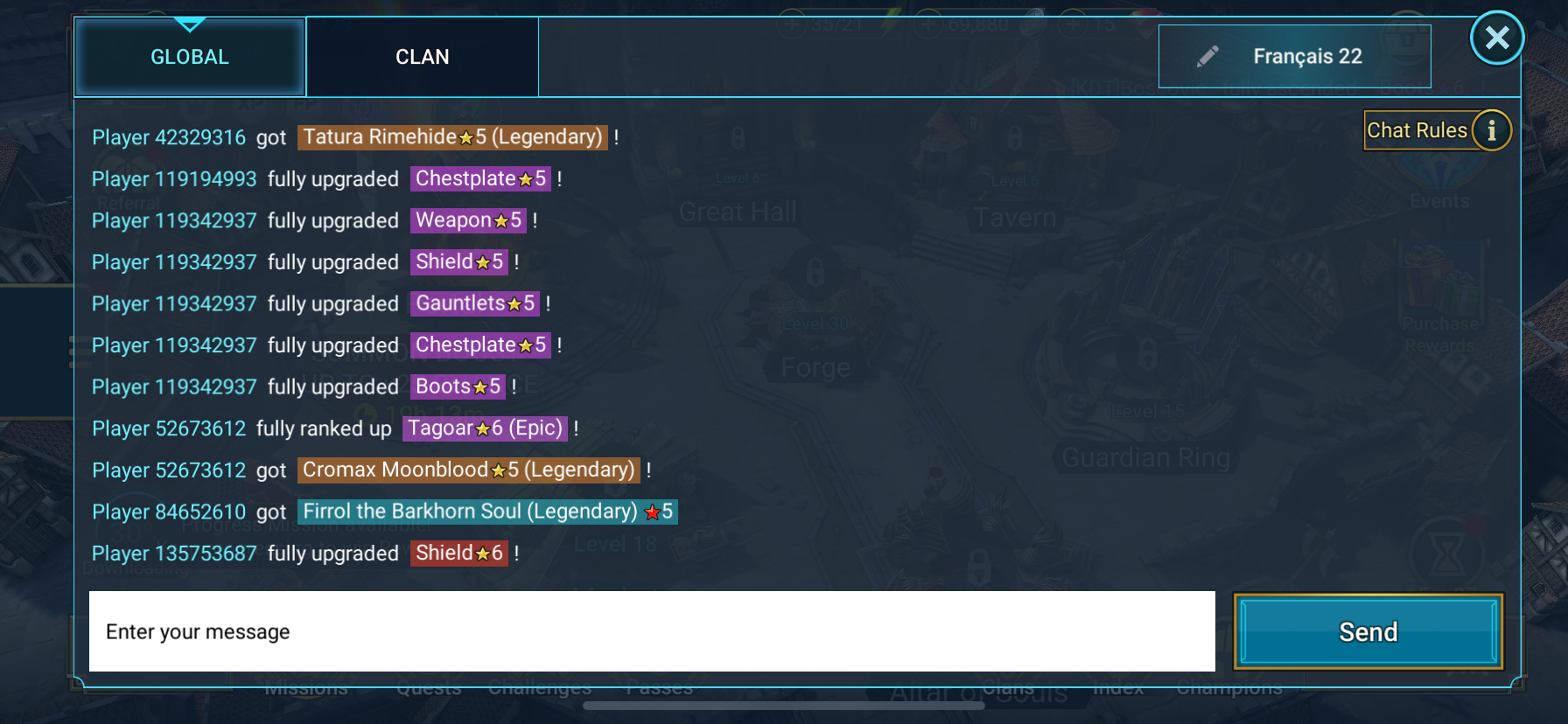}
    \label{fig:Raid Chat}
    \begin{flushleft} 
        \footnotesize
        \setlength{\parskip}{0em} 
       In the mobile game \textit{Raid: Shadow Legends}, a global chat exists that shares rare wins obtained by other gamers. Only rare and valuable rewards are reported.
    \end{flushleft}
\end{figure}